%% file: paper.tex
\documentclass[notitlepage,superscriptaddress,aps,pra,nofootinbib]{revtex4-1}
    \usepackage[bold]{hhtensor}
    \usepackage[pretty,uselistings]{revquantum}
        \newaffil{PNNL}{Pacific Northwest National Laboratory, Richland, Washington 99354, USA}

        \newnew{sc}{\textsc}
        \newsc{Prepare}
        \newsc{Select}
		\newsc{Had}
		\newsc{X}
		\newsc{Y}
		\newsc{Z}
		\newsc{Not}
		
        \newrm{ini}
        \newsc{Cnot}

    	\newcommand{\Phase}{\textsc{S}}
		\newcommand{\I}{\mathbb{1}}

		\newcommand{\hartree}{Hartree}

        \IfFileExists{sourcecodepro.sty}{
            \usepackage[semibold]{sourcecodepro}
        }{}

        \lstdefinelanguage{nwchem}{
            language = bash,
            morekeywords={basis,scf,end,geometry,tce,set,task,echo,start,memory,charge},
            commentstyle=\rmfamily\color{comment-color}
        }
        \lstdefinestyle{nwchem}{
            language=nwchem
        }
        \lstdefinestyle{csharp}{
            language=[Sharp]C,
            escapeinside={//*}{*//},
            morekeywords={partial, var, value, get, set},
            breaklines=true
        }

        \lstdefinelanguage{PowerShell}{}

	\newcommand\YAMLcolonstyle{\color{red}}
	\newcommand\YAMLkeystyle{\color{black}}
	\newcommand\YAMLvaluestyle{\color{black}}
	
	\makeatletter
	
	\newcommand\language@yaml{yaml}
	
	\expandafter\expandafter\expandafter\lstdefinelanguage
	\expandafter{\language@yaml}
	{
		basicstyle=\ttfamily\small\footnotesize,
		keywords={true,false,null,y,n},
		keywordstyle=\color{darkgray}\bfseries,
		sensitive=false,
		comment=[l]{\#},
		escapeinside={//*}{*//},
		stringstyle=\YAMLvaluestyle,
		moredelim=**[il][\YAMLcolonstyle{:}\YAMLvaluestyle]{:}   
	}
	
	\lst@AddToHook{EveryLine}{\ifx\lst@language\language@yaml\YAMLkeystyle\fi}
	\makeatother
	
	\usepackage{footnote}

     \usepackage{qsharp}
        \lstMakeShortInline[style=QSharp]"

    \usepackage{subfigure}

    \usepackage{complexity}
    \usepackage{subfiles}

	\usepackage{enumitem}
	
	\usepackage{tabularx}
	
	\usepackage{multirow}

	\usepackage{silence}
    \newcommand{\qsharp}{Q\#}
    \newcommand{\csharp}{C\#}
    \newcommand{\nwchem}{NWChem}
    
    \newcommand{\figurefolder}{../fig}
    
    \newcommand{\sourcefolder}{../src}

\renewcommand{\figurefolder}{.}
\renewcommand{\sourcefolder}{anc/src}

\begin{document}

\title{\qsharp~and \nwchem: Tools for Scalable Quantum Chemistry on Quantum Computers}
\date{\today}

\author{Guang Hao Low}
	\email{guanghao.low@microsoft.com}
	\affilMSRQuArC
\author{Nicholas P. Bauman}
	\affilPNNL
	
\author{Christopher E. Granade}
    \affilMSRQuArC

\author{Bo Peng}
    \affilPNNL

\author{Nathan Wiebe}
    \affilMSRQuArC

\author{Eric J. Bylaska}
	\affilPNNL

\author{Dave Wecker}
	\affilMSRQuArC

\author{Sriram Krishnamoorthy}
	\affilPNNL

\author{Martin Roetteler}
	\affilMSRQuArC

\author{Karol Kowalski}
	\affilPNNL
	
\author{Matthias Troyer}
	\affilMSRQuArC

\author{Nathan A.~Baker}
	\affilPNNL
	
\lstMakeShortInline[style=nwchem]@

\begin{abstract}
\subfile{\sectionfolder/abstract.tex}
\end{abstract} 

\maketitle
\makeatletter
\def\l@subsubsection#1#2{}
\makeatother
\tableofcontents
\newpage
\section{Introduction}
\label{sec:intro}

\subfile{\sectionfolder/intro.tex}

\section{Review}
\label{sec:review}
\subfile{\sectionfolder/review.tex}
\subsection{Quantum computing and programming}\label{sec:reviewQC}
\subfile{\sectionfolder/review_introqc.tex}

\subsubsection{Developing quantum algorithms in \qsharp}

\subfile{\sectionfolder/review_qsharp.tex}

\subsubsection{Running \qsharp~quantum algorithms}

\subfile{\sectionfolder/review_run_qsharp.tex}

\subsection{Quantum chemistry}
\label{sec:review_quantum_chemistry}

\subfile{\sectionfolder/review_qchem.tex}

\subfile{\sectionfolder/review_nwchem.tex}

\subsection{Quantum simulation}
\label{sec:review_of_quantum_simulation}

\subfile{\sectionfolder/review_simulation.tex}

\subsubsection{The Jordan--Wigner transformation}
\label{sec:review_jw}

\subfile{\sectionfolder/review_jordan_wigner.tex}
\subsubsection{Trotter--Suzuki simulation methods}
\label{sec:review_ts}

\subfile{\sectionfolder/review_trotter.tex}

\subsubsection{Qubitization simulation methods}
\label{sec:qubitization}

\subfile{\sectionfolder/review_qubitization.tex}

\subsubsection{Circuit optimizations for qubitization}
\label{sec:optimized_qubitization}
\subfile{\sectionfolder/review_qubitization_chem_optimized.tex}

\subsubsection{Phase estimation}
\label{sec:phase_estimation}
\subfile{\sectionfolder/algo.tex}

\section{The Broombridge schema for representing electronic structure problems}
\label{sec:representation}

\subfile{\sectionfolder/representation.tex}

\section{Simulating quantum chemistry with the Microsoft Quantum Development Kit}
\label{sec:simulation}

\subfile{\sectionfolder/simulation_introduction.tex}

\subsection{Constructing qubit Hamiltonians from chemistry Hamiltonians}
\label{sec:simulation_construct_hamiltonian}

\subfile{\sectionfolder/simulation_construct_hamiltonian.tex}

\subsection{Synthesizing quantum simulation circuits}
\label{sec:simulation_run_simulation}

\subfile{\sectionfolder/simulation_q_sharp.tex}

\subsection{Estimating eigenvalues}
\label{sec:simulation_estimate_energy}

\subfile{\sectionfolder/simulation_estimate_energy.tex}

\section{Example applications}
\label{sec:applications}

\subfile{\sectionfolder/applications.tex}

\section{Conclusions}
\label{sec:conclusions}
\subfile{\sectionfolder/conclusions.tex}
\acknowledgments{
We thank the broader Microsoft Quantum team for insightful comments and discussions.
The PNNL portion of this research was funded by the  Quantum Algorithms, Software, and Architectures (QUASAR) Initiative, conducted under the Laboratory Directed Research and Development Program at PNNL. 
KK and EJB also acknowledge  support from  the `Embedding Quantum Computing into Many-body Frameworks for Strongly Correlated  Molecular and Materials Systems' project, which is funded by the U.S. Department of Energy (DOE), Office of Science, Office of Basic Energy Sciences, the Division of Chemical Sciences, Geosciences, and Biosciences. 
The development of NWChem was supported from Environmental Molecular Sciences Laboratory (EMSL) operations. 
Environmental Molecular Sciences Laboratory (EMSL) operations are supported by the DOE Office of Biological and Environmental Research. 
The Pacific Northwest National Laboratory, is operated for the U.S. DOE by Battelle under Contract Number DE-AC05-76RL01830. 
}

\nocite{apsrev41Control}
\bibliographystyle{apsrev4-1}
\bibliography{apsrev-control,paper}

\appendix

\section{Running examples}
\label{sec:running-examples}

\subfile{\sectionfolder/apx-running-examples.tex}


\end{document}

%% file: abstract.tex

Fault-tolerant quantum computation promises to solve outstanding problems in quantum chemistry within the next decade. 
Realizing this promise requires scalable tools that allow users to translate descriptions of electronic structure problems to optimized quantum gate sequences executed on physical hardware, without requiring specialized quantum computing knowledge.
To this end, we present a quantum chemistry library, under the open-source MIT license, that implements and enables straightforward use of state-of-art quantum simulation algorithms.
The library is implemented in Q\#, a language designed to express quantum algorithms at scale, and interfaces with NWChem, a leading electronic structure package. 
We define a standardized schema for this interface, Broombridge, that describes second-quantized Hamiltonians, along with metadata required for effective quantum simulation, such as trial wavefunction ansatzes.
This schema is generated for arbitrary molecules by NWChem, conveniently accessible, for instance, through Docker containers and a recently developed web interface EMSL Arrows.
We illustrate use of the library with various examples, including ground- and excited-state calculations for LiH, H$_{10}$, and C$_{20}$ with an active-space simplification, and automatically obtain resource estimates for classically intractable examples.

%% file: intro.tex

Computational chemistry is one of the main consumers of computing resources today.
Computational chemistry calculations generally aim to approximate electronic structure Schr\"{o}dinger equation solutions to ``chemical accuracy'' (defined in Sec.~\ref{sec:reviewQC}): where predicted computed chemical properties quantitatively match experimental observations.
There have been great successes in the field, heralded by celebrated techniques such as density functional theory (DFT)~\cite{Wolfram2001dft}, coupled-cluster (CC) theory~\cite{Bartlett2007coupledcluster}, or density matrix renormalization group (DMRG)~\cite{Schollwock2005dmrg}.
However, chemical accuracy remains beyond the reach of tractable classical computing techniques for numerous problems, often involving transition metals and excited states.
In particular, a brute-force computational approach to the Schr\"{o}dinger equation has exponential cost arising from the curse of dimensionality, and is generally infeasible---for both current and projected supercomputers---for chemical systems beyond a hundred spin orbitals.

Quantum computing~\cite{Nielsen2004} promises a solution to this fundamental challenge of accurate electronic structure calculations.
Instead of simulating the time-evolution of electrons according to the laws of quantum mechanics on classical Turing-machine computers, quantum computers natively realize quantum effects at a hardware level.
The inherent computational power of quantum systems provides hope of solving the hardest quantum mechanical problems in chemistry and material science, such as the mechanism of biological nitrogen fixation~\cite{Reiher2016Reaction} or high-temperature superconductivity~\cite{Wecker2015Correlated,bauer2016hybrid}.

The \emph{theoretical} details of quantum algorithms for electronic structure calculations have been studied extensively. 
The first explicit algorithm for simulating generic local Hamiltonians was by Lloyd~\cite{Lloyd1996universal}, which has since seen continual improvements and generalizations~\cite{Aharonov2003Adiabatic, Berry2007Efficient, Berry2012, Berry2015Truncated, Low2016HamSim, Low2016Qubitization, Haah2018quantum, Low2018IntPicSim, Earl2018Random}. 
These algorithms have been specialized to fermionic systems~\cite{Abrams1997HubbardSimulation}, especially that of chemistry~\cite{whitfield2011simulation,Babbush2016Exponential}, along with numerous case studies~\cite{Wecker2014,poulin2014trotter,Hastings2015chemistry,babbush2015chemical,Childs2017Speedup,Babbush2018encoding}, and novel quantum-classical hybrid schemes that trade-off quantum circuit depth for at least polynomially more rounds of classical repetition and post-processing~\cite{bauer2016hybrid,McClean2016eigensolver,Rubin2018Marginals}.
A more thorough overview can be found in other publications~\cite{Cao2018ChemistryReview}.

However, the \emph{practical} details of using quantum methods for many real-world chemistry and material science problems pose unique challenges.
Setting aside the availability of fault-tolerant quantum hardware, and the difficulty of controlling said devices, it is non-trivial to program quantum devices to achieve a desired effect.
Paralleling the history of classical computing, quantum computing requires significant software development effort before domain experts can apply quantum resources to their problems at scale.
This need has motivated the recent development of a variety of quantum programming languages~\cite{green2013quipper,wecker2014liqui,steiger2016projectq,cross2017open,Svore2018qsharp}, each of which makes feasible and accessible various aspects of quantum software development.
Building on this, a number of different libraries for quantum chemistry applications have been developed~\cite{mcclean2017fermilib,mcclean2017openfermion}, in or for use with quantum programming frameworks, focusing primarily on near-term quantum chemistry tasks for Noisy Intermediate-Scale Quantum (NISQ)~\cite{Preskill2018NISQ} devices.

\begin{figure}[b]
	\includegraphics[scale=0.6,trim={0.1cm 0.5cm 0.5cm 1.0cm},clip]{\figurefolder/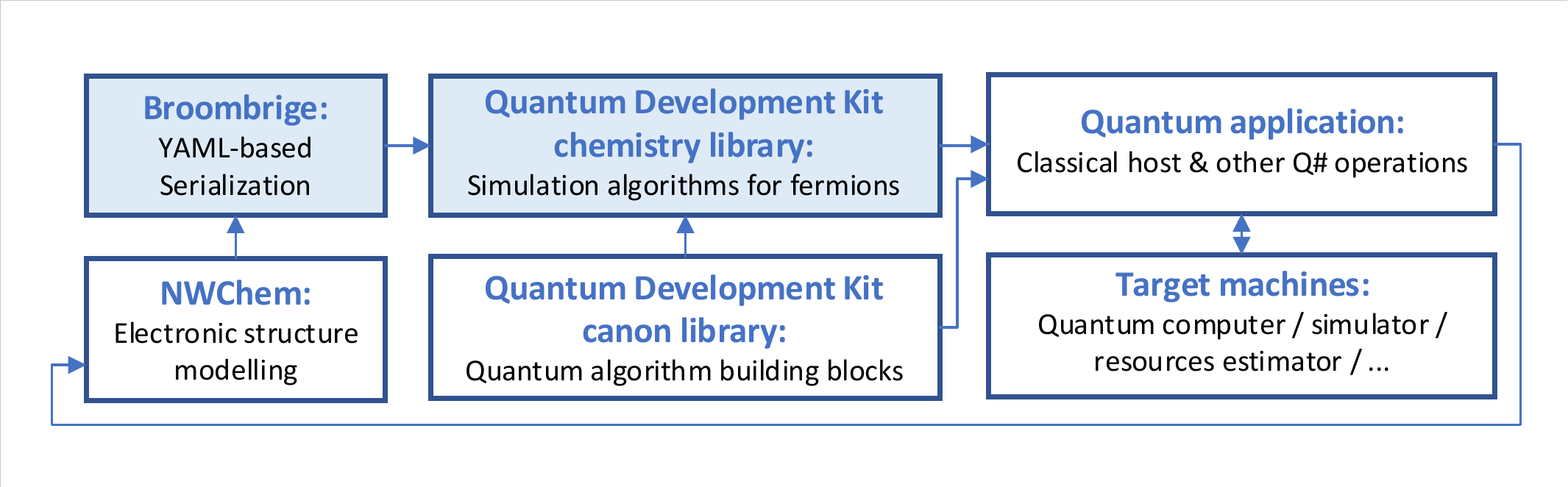}
	\caption{
		\label{fig:interplay_outline}
		Workflow for simulating quantum chemistry on quantum computers using NWChem and Microsoft Quantum Development Kit libraries.
		Our main contributions, Broombridge and the chemistry library, are shaded.
		See~\autoref{fig:interplay} for a detailed breakdown.
	}
\end{figure}

In this paper, we present a software suite outlined in~\autoref{fig:interplay_outline} that empowers quantum chemistry experts to write quantum simulation code that can be tested and costed using a classical computer as well as be executed as written on a fault-tolerant quantum computer once one becomes available. 
We accordingly focus on these future applications, as well as developing technologies that allow us to today simulate and profile resources needed for fault-tolerant quantum simulations.
The quantum simulation software we provide interacts with the underlying quantum error-correcting code and, in turn, the physical qubits through an abstraction that we call a simulator.
The simulator can be easily swapped with genuine hardware, guaranteeing that our code can be reused once fault-tolerant quantum hardware becomes available.
We have designed our quantum simulator software so that solutions are amenable to use by domain experts in quantum chemistry, without requiring strong domain expertise in quantum computing and quantum algorithms.
This focus is especially critical as we transition from preliminary investigations---such as the use of quantum devices to study ground-state energies of relatively simple molecules~\cite{Kandala2017variational}---to applications such as studying higher-energy properties and more complicated systems.

Making effective use of quantum computing resources past this transition puts a more significant demand on software development and interoperability between different pieces in a complete workflow.
We address this demand by introducing Broombridge, a new serialization format for representing fermionic Hamiltonians.
Broombridge enables interoperation between North-West Chemistry (NWChem) modelling software~\cite{valiev2010nwchem}, a leadership-class suite of tools for modeling quantum chemistry problems, and the Microsoft Quantum Development Kit~\cite{Svore2018qsharp}, a software platform for implementing quantum algorithms for both simulated execution and execution on eventual fault-tolerant hardware.
Our reproducible workflow shown in \autoref{fig:interplay_outline} automates simulations of large families of molecules. 
This workflow begins with NWChem, and its recent optional web interface Environmental Molecular Science Laboratory (EMSL) Arrows, which serializes descriptions of electronic structure problems as Broombridge.
Broombridge is then consumed by our quantum chemistry library in the Microsoft Quantum Development Kit, which is used in applications invoking quantum simulation and other supporting quantum algorithms.
As seen in the flowchart, many intermediate choices that affect performance and accuracy can be made between the initial problem specification and the final simulation on hardware. 
Ultimately, the software should free users from these fine details, and allow them to focus on the scientific endeavor.
\begin{figure}[b]
	\includegraphics[scale=0.6,trim={0.1cm 0.1cm 0.1cm 0.1cm},clip]{\figurefolder/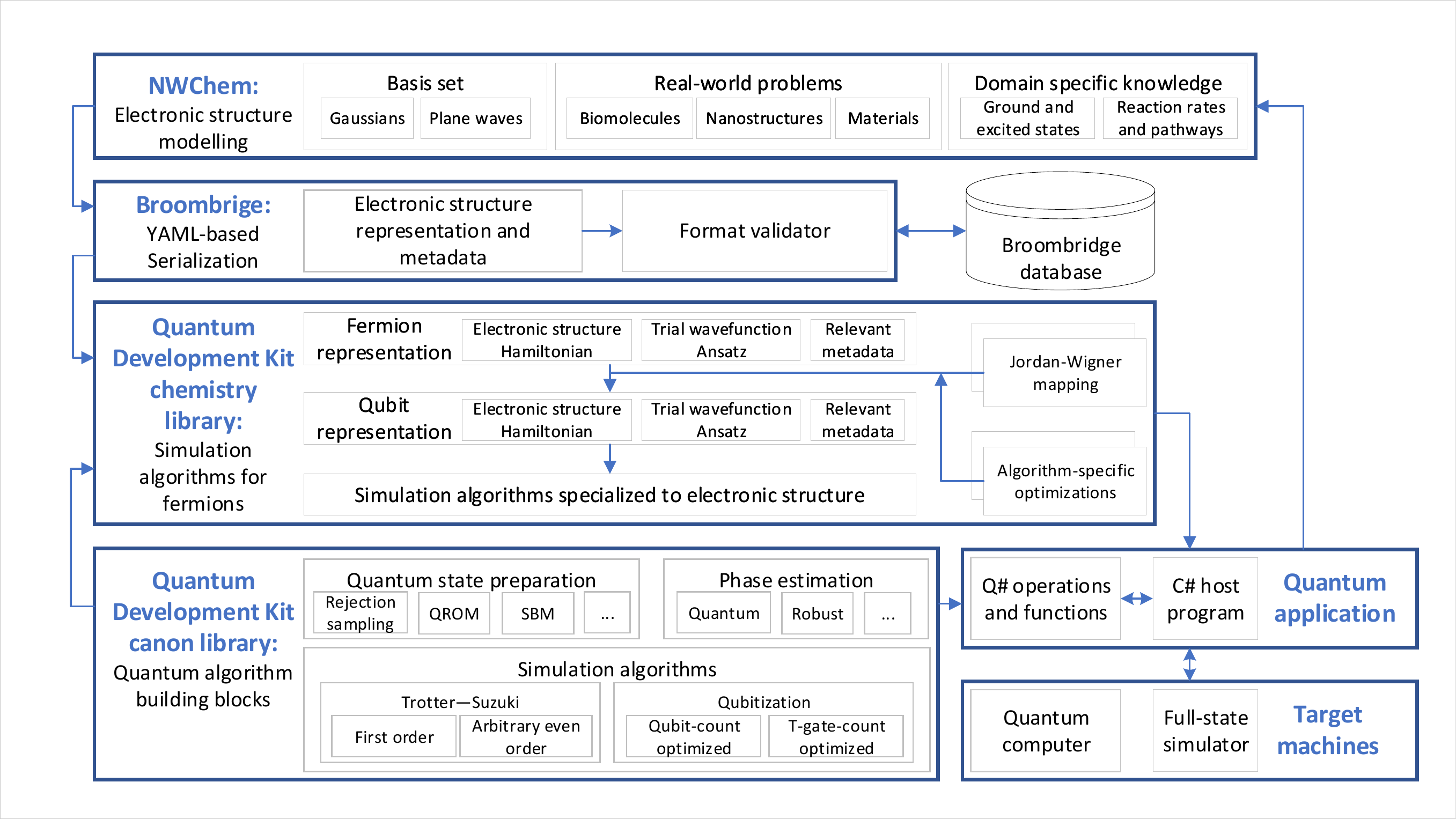}
	\caption{
		\label{fig:interplay}
		Detailed workflow for simulating quantum chemistry on quantum computers.
	}
\end{figure}

We illustrate this interoperability through a series of examples, including the basic examples traditionally used to introduce quantum chemistry development as well as examples motivated by future quantum simulation applications.
In particular, we highlight features in the Broombridge schema that enable us to conveniently describe the quantum Hamiltonian as well as the initial guesses for the eigenstate in question, which includes excited states that are traditionally difficult to probe by variational approaches.
These features are illustrated in example simulations of excited states of LiH and standard problems such as obtaining correlation energies of the hydrogen chain H$_{10}$, and energy calculations of different C$_{20}$ isomer configurations calculated using small active spaces. 
We highlight the ability of the quantum chemistry library to perform gate count estimates for a challenging example, specifically a full-configuration interaction simulation of C$_{20}$ in a $100$ spin-orbital active space.
Such simulations are beyond the reach of any classical computer, but are expected to be tractable for quantum computers.
These examples demonstrate the value of our system:  in addition to simplifying quantum resources counts and electronic structure simulations, it also enables large-scale surveys of quantum chemistry simulations that have hitherto been too challenging to perform by hand.

The layout of the paper is as follows.
We begin in~\autoref{sec:review} with a review of quantum computing as well as quantum chemistry.
In particular, we review state-of-the-art quantum simulation methods, such as qubitization and Trotter formulas, as well as methods for phase estimation.  
Important software such as \qsharp~and NWChem are also reviewed.  
In~\autoref{sec:representation}, we introduce the Broombridge schema used to interface NWChem with~\qsharp, and demonstrate how Broombridge may be produced.
\autoref{sec:simulation} provides a high-level discussion that shows how to use these tools together to simulate molecules within the Microsoft Quantum Development Kit chemistry simulation library.  
We provide concrete examples of the library in action in~\autoref{sec:applications} and use our tools to study the electronic states of LiH, H$_{10}$, and C$_{20}$ using quantum algorithms simulated on classical computers, as well as provide resource estimates for simulations of hard molecules before concluding.
Finally, we conclude in~\autoref{sec:conclusions} with our perspective of future directions for this line of work.

%% file: review.tex

The exponential growth of classical computing capabilities projected by Moore's law is coming to an end~\cite{Waldrop2016MooresLaw}. 
However, many computational problems of scientific and technological interest remain out of reach. 
In light of this, quantum computing has emerged---amongst various technologies---as the leading contender for continued progress due to its potential for realizing further exponential speedups, at least for certain specialized problems. 
The central idea is to develop a device that can, within arbitrarily small error, implement at least a universal set of quantum transformation on a quantum state.
This suffices to capture the computation power of practical quantum computing, encapsulated by the complexity class \BQP.

Of all the applications of quantum computing, the quantum simulation of physics, chemistry, and materials is envisioned to be the most transformative; one of the earliest useful areas for quantum advantage.
Since the dynamics of quantum systems are given by unitary transformations, we can in principle compile these dynamics, represented by Hamiltonians, into a sequence of discrete gates on a quantum computer.
This approach can yield exponential speedups over the best known classical algorithms for simulating hard quantum problems, such as those in catalysis or material science.

We review the key concepts of quantum computation based on qubits in~\autoref{sec:reviewQC}, together with an overview for implementing and execution quantum algorithms in the \qs~programming language. 
Subsequently, we state the fundamental concepts and definitions underlying quantum chemistry problems in~\autoref{sec:review_quantum_chemistry}, with a focus on its fermionic second-quantized representation, and the use of NWChem.
The ideas of quantum computation and quantum chemistry are merged in~\autoref{sec:review_of_quantum_simulation}, which outlines the map from fermions to qubit, and the algorithms that simulate quantum Hamiltonian on a quantum computer.

%% file: review_introqc.tex

Before describing the compilation of quantum simulation algorithm for Hamiltonian dynamics into primitive operations on quantum computing, we need to discuss the elementary units of quantum memory, and the target gate set of compilation.
The fundamental unit of memory in a quantum computer is a qubit.
A qubit is much like a probabilistic classical bit.
It can take the values $0$ or $1$, which we denote by the orthonormal two-dimensional column vectors $\ket{0}$ and $\ket{1}$.
The quantum state for a qubit can be an arbitrary quantum mixture of these two states; the simplest example is known as a ``pure'' quantum state.
For a qubit, the pure quantum state takes the form of a complex unit vector $a|0\rangle + b|1\rangle= \begin{bmatrix} a & b \end{bmatrix}^T$.
An ordinary probabilistic classical bit would have positive probabilities $a$ and $b$ that sum to $1$; however, a pure quantum state has $a,b \in \mathbb{C}$ and $|a|^2 + |b|^2=1$.

Just as measurement causes the prior distribution over the value of a classical stochastic bit to collapse to either $0$ or $1$, measurement of a quantum bit causes a similar impact on the quantum state.
The principal difference from the classical case is that $\Pr{[0]} = |a|^2$ for the quantum example, rather than $\Pr{[0]}=a$ as in the classical case. 
The exponentially greater power of quantum computers stems largely from this subtle difference.
As $a,b\in \mathbb{C}$ and---more importantly---can be negative, the different possible configurations that a register of qubits can be in can interfere with each other.
From this perspective, quantum computing can be viewed as the art of introducing and exploiting quantum interference for computational purposes.

The quantum state for multiple qubits can be represented by a tensor product of single qubit states. 
This means that, while a single qubit state is described by a two-dimensional vector space $\mathbb{C}^2$, a two-qubit state lives on a four-dimensional vector space $\mathbb{C}^{2^2}$.  
In general, an $n$-qubit quantum state exists on a vector space $\mathbb{C}^{2^n}$ of dimension $2^n$. 
This is unsurprising as a classical probability distribution over $n$ bits is also spanned by exponentially many bit strings.

Interference between the possible configurations of a quantum system is engineered using quantum gates.
Perhaps the most important defining characteristic of a quantum computer is the existence of a ``universal quantum gate set''.
This set consists of operations that can approximate, within arbitrarily small error, any transformation permitted by quantum mechanics on qubit states.
These legal transformations are represented as unitary matrices, which preserve the lengths of vectors, conserve the value of $|a|^2 + |b|^2$ for qubits, and hence conserve probability.
In principle, arbitrary single-qubit rotations and an entangling gate such as a controlled-\Not~ can be used to generate any unitary transformation.
If we take $|0\rangle = \begin{bmatrix} 1 & 0 \end{bmatrix}^T$ and $|1\rangle = \begin{bmatrix} 0 & 1 \end{bmatrix}^T$, then these single qubit rotations can be easily expressed as exponentials of the following single-qubit Pauli operations
\begin{align}
    \I = \begin{bmatrix} 1 & 0 \\ 0 &1 \end{bmatrix},\qquad \X = \begin{bmatrix} 0 & 1 \\ 1 &0 \end{bmatrix},\qquad \Y = \begin{bmatrix} 0 & -i \\ i &0 \end{bmatrix}, \;\text{and}\qquad \Z = \begin{bmatrix} 1 & 0 \\ 0 &-1 \end{bmatrix}.\label{eq:pauli}
\end{align}
These Paulis can be interconverted using products of Clifford gates called the Hadamard and phase gate:
\begin{align}
    \Had = \frac{1}{\sqrt{2}}\begin{bmatrix}1&1 \\ 1 & -1 \end{bmatrix},\qquad \Phase = \begin{bmatrix} 1& 0 \\ 0 & i\end{bmatrix}.
\end{align}
The notion of a single-qubit rotation is similarly a convenient concept in quantum computing.
The analogue of a rotation about the $z$-axis for a quantum bit for an angle $\theta$ can be expressed as 
\begin{align}
    R_\Z(\theta)= e^{-i \Z \theta/2}= 
    \cos{(\theta/2)}\I-i\sin{(\theta/2)}\Z=\begin{bmatrix} \cos{(\theta/2)} & -i\sin{(\theta/2)} \\ -i\sin{(\theta/2)} &\cos{(\theta/2)} \end{bmatrix}.
\end{align}
Any-single qubit operation can be expressed as a sequence of three rotations: $R_\Z(\theta) \cdot R_\X(\phi) \cdot R_\Z(\psi)=R_\Z(\theta)\cdot \Had \cdot R_\Z(\phi)\cdot \Had\cdot R_\Z(\psi)$ for appropriate Euler angles $\theta$, $\phi$ and $\psi$.
In fault-tolerant applications, these rotations are typically approximated by sequences of $\Had$ and $\T = \sqrt{\Phase}$ gates.

The simplest two-qubit quantum gate is the controlled-\Not~gate controlled by qubit $j$ and applied to qubit $k$, which has the action $\ket{0}_j\ket{x}_k \mapsto \ket{0}_j \ket{x}_k$ and $\ket{1}_j\ket{x}_k \mapsto \ket{1}_j \ket{x \oplus 1}_k$.
The gate takes the following matrix representation (using the above basis convention)
\begin{equation}
    \Cnot_{jk}= \begin{bmatrix} 1 &0 &0 &0 \\ 0&1 &0 &0 \\ 0&0&0&1 \\ 0&0&1&0 \end{bmatrix}.
\end{equation}
It can be useful to observe that gates in the Clifford group $\{\Had,\Phase,\Cnot\}$ can map any $n$-qubit Pauli $\P\in\{\I,\X,\Y,\Z\}^{\otimes n}$ operator to any other $n$-qubit Pauli by conjugation.
This provides one possible, though not necessarily the most efficient, implementation of $R_\P(\theta)= e^{-i \P \theta/2}$ which is commonly found in quantum simulation algorithms.

Thinking about quantum simulation algorithms strictly in terms of these operations is quite taxing, just as programming a word processor using only assembly code would be a challenge in classical computing.
Higher-level quantum programming languages have been developed to address these challenges~\cite{green2013quipper,wecker2014liqui,steiger2016projectq,Svore2018qsharp}.
These languages bridge the gap between the low-level physics-inspired description of the quantum states and the higher level descriptions of algorithms that are customarily shown as pseudocode in quantum computing papers.
Such bridges are essential not only because the act of compiling a quantum algorithm into an optimized sequence of quantum gates is demanding, but also because a quantum computer is not just a single monolithic device.
A quantum computer is a rich nested stack of computing substrates that view the quantum computer at different levels of abstraction:  a fault-tolerant quantum computer provides a user with a view of logical qubits that are made out of collections of physical qubits which are themselves an abstraction of the basic physical systems that lie beneath all held within an error correcting code wherein the quantum gates actually represent complex sequences of physical gates.
Given this complexity, one quickly realizes that high-level quantum programming languages are not a luxury, but a necessity---even before quantum computing comes of age.

%% file: review_qsharp.tex

Our aim is to show how simulations of quantum chemistry can be made easier by using the Microsoft Quantum Development Kit in conjunction with NWChem.
Here we will review the Microsoft Quantum Development Kit which provides a new language, \qs, that is used to program the quantum chemistry simulations in this paper.
The Microsoft Quantum Development Kit is distributed under the open-source MIT license as a set of installable packages for .NET Core, an open-source cross-platform programming environment that includes high-level classical languages such as C\# and F\#.

\begin{figure}
    	\includegraphics[scale=0.60,trim={0.1cm 0.1cm 0.1cm 0.1cm},clip]{\figurefolder/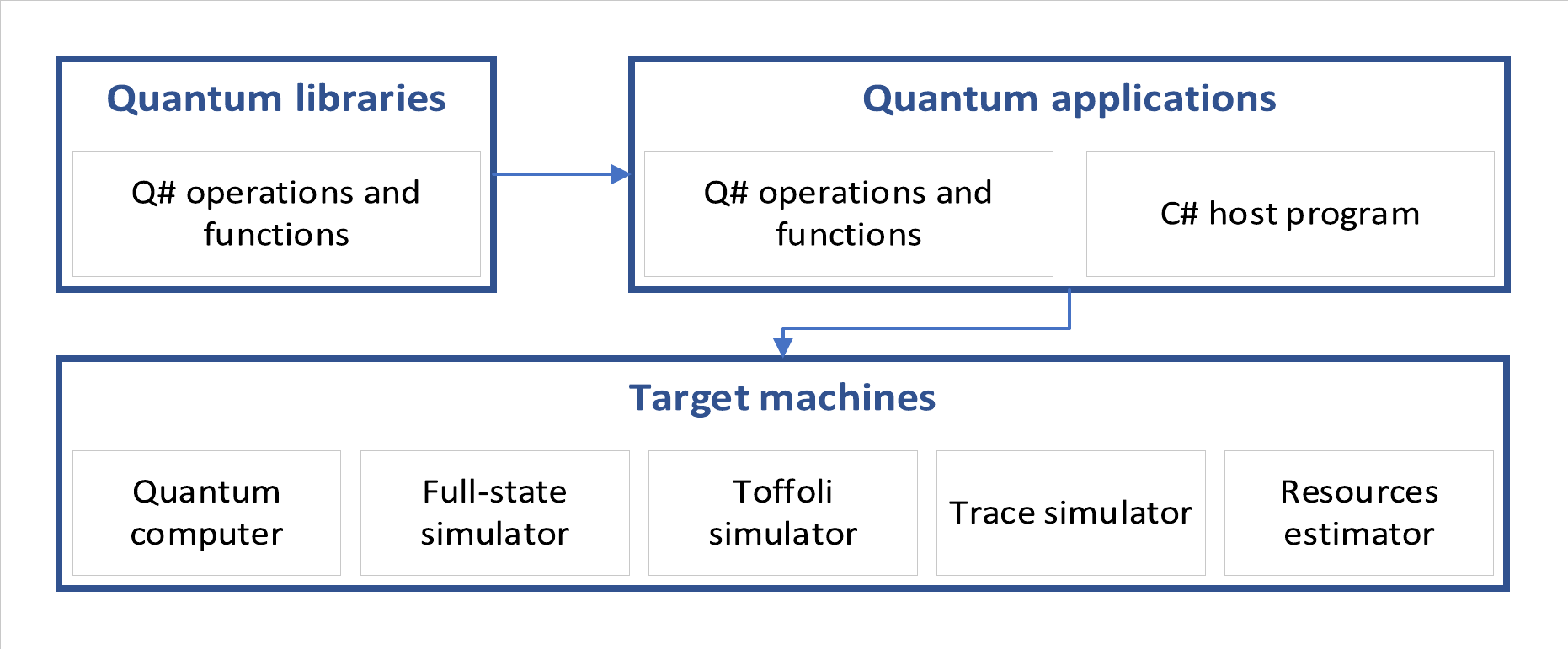}
    \caption{
        \label{fig:qdk-stack}
        Execution model used by the Microsoft Quantum Development Kit to interact with \emph{target machines}, each of which executes \qs~code on behalf of a classical program written in a .NET Core language such as C\#.
    }
\end{figure}

Quantum programs written in \qs~use an accelerator model, similar to graphics programming or the use of field-programmable gate arrays.
As illustrated in \autoref{fig:qdk-stack}, once a user writes \qs~code, that program can be dispatched to one of several \emph{target machines} by a host program written in any .NET Core language.
The target machine then runs the \qs~program, including both classical logic and quantum instructions, and returns its result to the classical host program.
Each target machine exposes a set of available instructions to \qs~programs as \emph{operations} that can be called during a program's execution.
For instance, the Hadamard gate can be applied by calling the "H" operation in the "Microsoft.Quantum.Primitive" namespace.
The operations that define the interface between a \qs~program and target machines are collectively known as the \emph{prelude}, and can be referenced in a \qs~source file using an "open" statement.

In \autoref{lst:qrng}, we show a simple example of using the Microsoft Quantum Development Kit to program a quantum random number generator (QRNG).
At \autoref{lst:qrng:declaration}, the user declares a new operation, "Qrng", that can interact with the target machine in a variety of ways, including allocating fresh qubits, or calling primitive operations.
Once defined in this way, "Qrng" can be called from other operations, or can be invoked from a classical host program written in C\#~or another .NET Core language.
For a complete set of samples demonstrating this process, please see~\url{https://github.com/Microsoft/Quantum}.
\begin{lstlisting}[
style=QSharp,
gobble=0,
label={lst:qrng},
firstline=4,
caption={A simple quantum random number generator (QRNG) written in \qs.\hfill\texttt{examples/qrng/qrng.qs}}
]
// All Q# code lives inside of a namespace, so we begin by declaring that namespace here.
namespace Qrng {
    // We can import names from other namespaces using the //*open*// keyword.
    open Microsoft.Quantum.Primitive;

    /// # Summary
    /// Prepares a qubit in the $\ket{+}$ state, then measures in the $\Z$ basis to obtain a random bit.
    // This line declares an operation that takes no inputs and returns a Result.
    operation Qrng() : Result { //* \label{lst:qrng:declaration} *//
        mutable result = Zero;
        // The //*using*// keyword asks the target machine for a fresh qubit, initially in the $\ket{0}$ state.
        using (qubit = Qubit()) {
            // Applying the //*H*// operation to a qubit in the $\ket{0}$ state prepares that qubit in the $H\ket{0} = \ket{+}$ state.
            H(qubit);
            // We can measure our qubit in the $\Z$-basis using the //*M*// operation provided by the prelude.
            set result = M(qubit);
            // Before returning the qubit to the target machine, we need to make sure that it is in the $\ket{0}$ state. This can be done using a classically controlled $\X$ gate, which we apply using a classical //*if*// statement and the //*X*// operation.
            if (result == One) { X(qubit); } //* \label{lst:qrng:reset} *//
        }
        // Finally, we return our random bit to the classical host program.
        return result;
    }
}
\end{lstlisting}

The Microsoft Quantum Development Kit also includes a set of standard libraries known as the \emph{canon} that are built up within \qs~itself.
These libraries provide \qs~programs with useful routines for performing measurements and manipulating flow control, as well as for higher-level quantum algorithms such as the quantum Fourier transform (QFT), implementations of phase estimation algorithms, and routines for quantum simulation algorithms.
A complete reference to the prelude and canon is available online at \url{https://docs.microsoft.com/qsharp/api/}.
The functions and operations in the canon, together with other features of the \qs~language---such as the "Adjoint" \emph{functor} to automatically transform an operation into its inverse operation, make it straightforward to encapsulate and reuse code in quantum applications.

%% file: review_run_qsharp.tex
\lstMakeShortInline[style=csharp]!

Once a quantum program has been written using \qsharp, it can be run using a classical host program to allocate a target machine.
This host will often be written in \csharp, but any other .NET Core language can be used.
To demonstrate, we use an excerpt in~\autoref{lst:teleport} from the example of quantum teleportation, which is described in detail within the complete source file.
\begin{lstlisting}[
style=QSharp,
gobble=0,
linerange={49-68},
label={lst:teleport},
caption={An example of quantum teleportation between two qubits written in \qsharp.\hfill\texttt{examples/teleport/Teleport.qs}}
]
    /// # Summary
    /// Runs the Teleport operation above by allocating three qubits,
    /// and asserting that the message qubit is correctly teleported.
    operation RunTeleport() : Unit {
        using ((msg, here, there) = (Qubit(), Qubit(), Qubit())) {
            // We'll prepare a state $\ket{+} = H\ket{0}$ by using
            // the //*H*// operation.
            H(msg);
            // After teleporting using the //*Teleport*// operation, 
            // the target qubit should be in the $\ket{+}$ state. Thus, 
            // applying another //*H*// will transform it back into the 
            // $\ket{0}$ state.
            Teleport(msg, here, there);
            H(there);
            // When run on a simulator, we can assert this fact
            // using the AssertQubit operation. In other cases, this
            // operation is safely ignored since it has no effect.
            AssertQubit(Zero, there);
        }
    }
\end{lstlisting}
In many cases, we want to run our quantum programs on a simulator that will let us check that they operate correctly on noiseless qubits.
The Microsoft Quantum Development Kit provides the !QuantumSimulator! target machine for this case, as demonstrated in \autoref{lst:teleport-host}.
Full details on running this example can be found in \autoref{sec:running-examples} of the appendix.

The !QuantumSimulator! target machine is especially useful in conjunction with unit testing frameworks such as xUnit~\cite{meszaros2007xunit}, as this makes it possible to write a comprehensive set of tests for a quantum algorithm implementation.
Such test suites help build confidence that an implementation is correct.

Once we are confident that a \qsharp~program functions correctly, the next steps often involve costing out larger cases that are intractable with only classical resources.
The Microsoft Quantum Development Kit offers another target machine, the !QCTraceSimulator! class, which counts the resources required to run a quantum program.
We demonstrate the use of this target machine in \autoref{lst:teleport-host} as well.
\lstinputlisting[
	style=csharp,
	gobble=0,
	linerange=main-end,
	label={lst:teleport-host},
	caption={Classical host for \autoref{lst:teleport}.\hfill\texttt{examples/teleport/Host.cs}}
]{\sourcefolder/examples/teleport/Host.cs}

Critically, the \qsharp~code run in both parts of \autoref{lst:teleport-host} is identical: the "RunTeleport" operation cannot observe whether it is is being simulated by !QuantumSimulator! or !QCTraceSimulator!.
That is, the choice of target machine is \emph{transparent} to \qsharp~code as it is being run, allowing us to build confidence by testing with small classical resources, and then reuse the same code in cost estimation and---eventually---in actual hardware.

\lstDeleteShortInline!

%% file: review_qchem.tex

The main focus of quantum chemistry is providing computational tools for modeling molecular structure, chemical reactions, dynamics, and spectroscopic properties.
These are inextricably linked to many-body methods for solving the stationary (time-independent) Schr\"odinger equation
\begin{equation}
    H\ket{\Psi} = E \ket{\Psi} \;,
\label{hpsi}
\end{equation}
where $H$, $\ket{\Psi}$, and $E$ represent the Hamiltonian operator, wavefunction, and corresponding energy of a molecular system (respectively). 
The energy scale of interest in these problems is typically on the order of $1$~\hartree, which we shall use as our units for energy and inverse-time in the following.
In general, eigenvectors and eigenvalues of the Schr\"odinger equations ($\ket{\Psi}$ and $E$, respectively) describe ground or excited electronic states.
By applying various assumptions regarding the nature of the inter-electron interactions and the algebraic form of the electronic wavefunction, a plethora of various approximate methods have been introduced and tested over the last century.
These methods find approximate solutions to Schr\"odinger equations, and their underlying assumptions intrinsically define the memory requirements and numerical overheads of simulating approximate many-body models on classical machines. 

Among the several classes of many-body methodologies, a number of approaches stand out. 
These include the numerous variants of Hartree--Fock (HF) and DFT methods, many-body perturbation theory (MBPT), Green's function methods (GF), configuration interaction (CI) and CC methods, density matrix theory, and DMRG approaches.
Over the years, these methods have evolved into staple working engines used in numerous simulations of chemical processes. 
A significant effort has also been directed towards the development of reduced scaling methods and embedding formulations to handle correlation effects in  large molecular systems. 
The widespread use of these methods has emphasized the role played by the proper inclusion of complex electron correlation effects for a comprehensive and accurate understanding of molecular processes. 
In some cases, for example CI and  CC theories, achieving ``chemical accuracy'' of roughly
\begin{align}
    \text{Chemical accuracy}
    =10^{-3}\;\text{\hartree}
    =0.02721\;\text{eV}
    =2.625\;\text{kJ/mol}
    =316.8\;k_B\text{Kelvin},
\end{align} 
requires including enormous numbers of wavefunction parameters, which results in a steep computational complexity scaling of these formalisms.
In asymptotic limit of the Full CI (FCI) formulation, the number of wavefunction parameters scales as $N!$ with respect to system size $N$.
Examples that appear to require reaching this limit include modeling low-spin open-shell systems, radicals, transition metal oxides, and actinides. 
Fortunately, quantum computing, which has polynomial scaling in $N$, offers means to address the exponential scaling of high-accuracy wavefunction formulations on classical computing platforms.

\subsubsection{Second-quantized Hamiltonians}
\label{sec:second-quantized}

The language of second quantization has permeated almost the entire area of quantum chemistry and it is widely used to classify various many-body effects contributing to complex inter-electron correlation effects.
Second quantization has also become a foundation for diagrammatic representation of various many-body theories.
The central role in quantum chemistry is played by the Born--Oppenheimer approximation, where the electronic structure Hamiltonian $H$ describes electrons that move within a fixed nuclear frame
\begin{align}
    \label{eq:firstq}
    H &= \sum_{i=1}^{\eta}\left(-\frac{\hat{{\bf\nabla}}_i^2}{2} - \sum_{l} \frac{Z_l}{|{\bf r}_i-{\bf R}_l|}\right)+\sum^\eta_{i<j} \frac{1}{|{\bf r}_i-{\bf r}_j|},
\end{align}
where $\hat{{\bf \nabla}}^2$ is the Laplacian, $Z_l$ and ${\bf R}_l$ are the charge and position of the $l$-th nucleus, and ${\bf r}_i$ is the position of the $i$-th electron. In this basis, the corresponding $\eta$-electron wavefunction has components $\Psi({\bf x}_1,{\bf x}_2,\ldots,{\bf x}_\eta)=\braket{{\bf x}_1{\bf x}_2\ldots{\bf x}_\eta|\Psi}$ indexed by the spatial and spin coordinates of $i$-th electron, i.e., ${\bf x}_i=({\bf r}_i,s_i)$ where ${\bf r}_i\in\mathbb{R}^3$ and $s_i\in\{\uparrow,\downarrow\}$.

As $\ket{\Psi}$ describes of indistinguishable fermions (electrons), it also has to satisfy fermion statistics associated with its antisymmetry upon swapping the coordinates of any pair of electrons.
Second-quantization techniques and the occupation number representation provide a concise way of characterizing many-body effects in the Hamiltonian operator and electronic wavefunction while automatically assuring its anti-symmetry.
In the second-quantization, all operators and the many-body wavefunction are represented in terms of creation and annihilation operators $a_p^{\dagger}$ and $a_p$ indexed by $p$. These operators satisfy the of anti-commutation relations
\begin{equation}
    \lbrace a_p, a_q\rbrace=\lbrace a_p^{\dagger},a_q^{\dagger}\rbrace =0,\;\text{and}\quad\lbrace a_p,a_q^{\dagger}\rbrace = \delta_{pq}.
\label{anticom}
\end{equation}
Additionally, when acting with any annihilation operator on the physical vacuum state $\ket{0}$ or when the creation operator is applied to an occupied state, the following is always satisfied
\begin{equation}
\forall{p},\quad    a_p \ket{0} = a^\dagger_p \ket{1}_p = 0.
    \label{eq:ap0}
\end{equation}
Similarly, if the annihilation operator is applied to an occupied state or if the creation operator is applied to the vacuum then particles are destroyed or created respectively:
\begin{equation}
\forall{p},\quad a_p^\dagger \ket{0} = \ket{1}_p, \quad a_p \ket{1}_p = \ket{0}_p.
    \label{eq:ap1}
\end{equation}
Using these operators, one can represent the Hamiltonian operator as
\begin{equation}
    H=\sum_{pq} h_{pq} a_p^{\dagger}a_q + \frac{1}{2}\sum_{p,q,r,s}
    h_{pqrs}
    a_p^{\dagger} a_q^{\dagger}  a_r a_s,
    \label{eq:hsecond}
\end{equation}

The connection to the electronic structure Hamiltonian of~\autoref{eq:firstq} is completed by choosing $p=(i,s)\in\{1,\cdots,M\}\times\{\uparrow,\downarrow\}$ to index one of $M$ carefully chosen orbitals with amplitude $\phi_i({\bf r})=\braket{{\bf r}|\phi_i}$ in the position basis, and a spin state with amplitude $\chi_s$.
In other words, $a^\dagger_p$ creates an electron in one of $N=2M$ single-particle spin-orbitals $\ket{\psi_p}=\ket{\phi_i}\ket{\chi_s}$.
The coefficients of~\autoref{eq:hsecond} are then
\begin{align}
    \label{eq:h_pq}h_{pq}&\equiv h_{(i,s_1)(j,s_2)}= \bra{\psi_p}\left(-\frac{{\bf \nabla}^2}{2}- \sum_{l} \frac{Z_l}{|{\bf r}-{\bf R}_I|}\right)\ket{\psi_q}
    =\delta_{s_1s_2}\underbrace{\int d{\bf r} \;\phi_i^{\ast}({\bf r}) 
    \left(-\frac{{\bf \nabla}^2}{2}- \sum_{l} \frac{Z_l}{|{\bf r}-{\bf R}_I|}\right){\phi}_j({\bf r})}_{h_{ij}},  \\
    \label{eq:h_pqrs}h_{pqrs} &\equiv h_{(i,s_1)(j,s_2)(k,s_3)(l,s_4)}=
    \bra{\psi_p}\bra{\psi_q}\frac{1}{|{\bf r}_1-{\bf r}_2|} \ket{\psi_r}\ket{\psi_s}
    =\delta_{s_1s_4}\delta_{s_2s_3}
    \underbrace{\int d{\bf r}_1 d{\bf r}_2 
    \frac{\phi_i^{\ast} ({\bf r}_1) \phi_j^{\ast}({\bf r}_2)\phi_k ({\bf r}_2) \phi_l({\bf r}_1)}{|{\bf r}_1-{\bf r}_2|}}_{h_{ijkl}}.
\end{align}
A convenient way of representing electronic wavefunction vector $|\Psi\rangle$ is then as as a linear combination of all symmetry-allowed Slater determinants $\ket{f}$ created by some sequence of $\eta$ creation operators
\begin{equation}
    |\Psi\rangle = \sum_{f} c_{f} \ket{f}, 
    \label{fciex}
\end{equation}
where the $c_{f}$ are complex coefficients. 
The above expansion is often referred to as the full configuration expansion (FCI) and is considered an exact solution to the electronic Schr\"odinger equation for a given finite basis set. 
A given Slater determinant $|f\rangle$ in the occupation number representation can be expressed in terms of string of $f_i$ numbers ($f_i\in\lbrace 0,1 \rbrace$, $i=1,\ldots,N$, where $N$ is a total number of spin-orbitals), which is usually denoted as 
\begin{equation}
    |f\rangle = |f_{N},\ldots,f_i,\ldots,f_1\rangle \;,
    \label{fslater}
\end{equation}
where the actions of the creation and annihilation operators on such a state are given by formulas
\begin{widetext}
    \begin{eqnarray}
        a_p |f_{N},\ldots,f_i,\ldots,f_1\rangle &=& 
        \delta_{f_p,1} (-1)^{\sum_{i=0}^{p-1} f_i} 
        |f_{N},\ldots,f_i-1,\ldots,f_1\rangle  \;, \label{apf} \\
        \text{and}\;a_p^{\dagger}  |f_{N},\ldots,f_i,\ldots,f_1\rangle &=& 
        \delta_{f_p,0} (-1)^{\sum_{i=0}^{p-1}f_i} 
        |f_{N},\ldots,f_i+1,\ldots,f_1\rangle  \;, \label{apcf}
    \end{eqnarray}
\end{widetext}
Since the cost of solving FCI problem grows exponentially with the basis set size $M$, classical computers can only be used to solve small problems. 
In contrast, quantum computers can efficiently create and manipulate states that are a superposition of exponentially many elements.
Thus problems that are intrinsically multi-configurational and require FCI to achieve chemical accuracy appear to be ideal targets for quantum solutions.

%% file: review_nwchem.tex

\subsubsection{Computational Quantum Chemistry in NWChem}
\label{sec:review_nwchem}
The NWChem modeling software, found at~\url{http://www.nwchem-sw.org}, is a popular molecular chemistry simulation tool designed from conception to operate on massively parallel supercomputers~\cite{bernholdt1995parallel,kendall2000high,valiev2010nwchem}, and is open-source under the permissive Educational Community License (ECL) 2.0 license.
Source files and binaries for NWChem are located in a Github repository~\url{https://github.com/nwchemgit/nwchem}.
A Docker image of NWChem is also available~\url{https://hub.docker.com/r/nwchemorg/nwchem-qc}.
While prior NWChem releases are compatible with Linux distributions, this option is not currently recommended as the most recent versions of NWChem (compatible with Windows) are required to generate input for the Microsoft Quantum Development Kit.

Today, NWChem contains an umbrella of modules that include single and multi-configuration self consistent field (SCF); second-order M{\o}ller-Plesset perturbation theory; CC; selected CI; Tensor Contraction Engine (TCE) based many body methods; DFT; time-dependent DFT (TDDFT); real-time TDDFT; pseudopotential plane-wave DFT; band structure; {\it ab initio} molecular dynamics; Car--Parrinello molecular dynamics; classical molecular dynamics; QM/MM; AIMD/MM; GIAO NMR; COSMO, COSMO-SMD, and RISM solvation models; free energy simulations; reaction path optimization; parallel-in-time dynamics; among other capabilities.
New capabilities continue to be added with each release.

An electronic structure problem can be input to NWChem by specifying the coordinates of its component atoms, as seen in the following~\autoref{lst:nwchem_geometry} for a minimal example of Lithium Hydride (LiH).
\begin{lstlisting}[
style=nwchem,label={lst:nwchem_geometry},caption={NWChem input example for geometry of LiH electronic structure problem.\hfill\texttt{examples/lih.nw}},
linerange={1-23}
]
# Begins a calculation that will output a file named `lih.out'.
start lih

# Outputs a copy of this input deck in a file named `lih.nw'.
echo

# Allocates the maximum memory used for this calculation.
memory 1900 mb

# Overall charge of this system in units of electric charge.
charge 0

# Geometry specification input group of the electronic structure problem. 
# Atom positions are defined in Cartesian coordinates in units of Angstrom (//*{\color{comment-color}$10^{-10}$m}*//).
geometry units angstrom
# Specifying the point group symmetry of the problem can speed up calculations. Currently, only the //*{\color{comment-color}$C_{1}$}*// symmetry is allowed in the NWChem-Quantum Development Kit interface. 
symmetry c1
# A Lithium atom is placed at coordinates //*{\color{comment-color}$(0,0,0)$\AA}*//.
# A Hydrogen atom is placed at coordinates //*{\color{comment-color}$(0,0,1.624)$\AA}*//.
Li   0.000000000000000   0.000000000000000   0.000000000000000    
H    0.000000000000000   0.000000000000000   1.624000000000000
# The keyword `end' marks the end of an input group. 
end
\end{lstlisting}

Electronic structure Hamiltonians in the second-quantized formalism employ certain representations of molecular orbitals, usually corresponding to some independent particle model (IPM) based on a finite-dimensional one-particle basis set.
NWChem offers a broad array of IPMs including:
\begin{itemize}
	\item Restricted Hartree--Fock formalism (RHF),
	\item Open-shell  Restricted Hartree--Fock method (ROHF),
	\item Unrestricted Hartree--Fock method (UHF),
	\item various Density Functional Theory (DFT) formulations.
\end{itemize}
These methods can use a variety of basis sets---ranging from Gaussian to plane-wave---to express molecular orbitals $\ket{\phi_i}$ as linear combinations of other basis set orbitals $\ket{\Phi_{\mu}}$, i.e.,
\begin{equation}
	\ket{\psi_i} = \sum_{\mu} c_{i\mu} \ket{\Phi_{\mu}},
	\label{eq:lcao}
\end{equation}
where $c_{i\mu}$ are variationally optimized coefficients.
Once the molecular orbitals are determined, one- and two-electron integrals are obtained from atomic one- ($h_{\mu\nu}$) and two-electron ($h_{\mu\nu\rho\sigma}$) integrals through the so-called 2- and 4-index transformations:
\begin{align}
	h_{ij}= \sum_{\mu\nu} c_{i\mu} c_{i\nu} h_{\mu\nu}, \quad \text{and}\quad h_{ijkl} = \sum_{\mu\nu\rho\sigma} c_{i\mu} c_{j\nu} c_{k\rho} c_{l\sigma} h_{\mu\nu\rho\sigma}. \label{eq:2index}
\end{align}
In practical implementations, these transformations are factorized using recursive intermediate techniques with classical time complexity $\mathcal{O}(N^3)$ and $\mathcal{O}(N^5)$ respectively. 

Of the IPMs, DFT can provide an array of trial wavefunctions depending on the functional used to describe the system which bring a definite level of uncertainty and are not systematically improvable.
While the single-configuration HF state provides a zeroth-order approximation to the ground state, its description of the trial wavefunction is often qualitatively poor or incorrect and provides energy estimates that far exceed chemical accuracy, often by orders of magnitude.
In order to provide better target trial wavefunctions one can turn to post-HF approximations which aim to recover the difference between the IPM and FCI, such as MBPT, CI formalisms, and CC methodologies.

Higher-level methods rely on an initial calculation based on the choice of basis set and IPM.
For example, a calculation with RHF orbitals in the Slater-type orbital (STO)-3G basis set is made through two groups of instructions in the NWChem input as seen in~\autoref{lst:nwchem_basis_and_IPM}.
A similar input structure may be used to produce DFT orbitals in various basis sets.
These molecular orbitals in~\autoref{eq:lcao} are subsequently used in~\autoref{eq:2index} to generate one- and two-electron integrals defined by~\autoref{eq:h_pq} and~\autoref{eq:h_pqrs}.
\begin{lstlisting}[
style=nwchem,label={lst:nwchem_basis_and_IPM},caption={NWChem input instructions for choice of basis set and independent particle model.\hfill\texttt{examples/lih.nw}},
linerange={25-46}
]
# Basis set specification input group.
basis	
# We choose the STO-3G basis set for both atoms.
H library sto-3g
Li library sto-3g
# End of basis set input group.
end

# Self-consistent field model specification input group.
scf	
# We choose the restricted Hartree--Fock approach.
rhf	
# We choose a system where all electrons are paired if possible.
singlet	
# SCF convergence threshold for molecular orbital variational optimization in //*\autoref{eq:lcao}*//.
thresh 1.0e-10	
# Two-electron integral screening threshold for the evaluation of the energy and related quantities.
tol2e 1.0e-10	
# Maximum number of iterations in variational optimization to reach thresholds.
maxiter 501
# End of SCF model input group.
end
\end{lstlisting}

Of the post-IPM methods, CC theory provides a rapid convergence to the FCI limit and is systematically improvable along with other desirable features.
The most common IPM for CC calculations is the HF wavefunction, $\ket{\Psi_\mathrm{HF}}$.
The ground-state CC wavefunction $\ket{\Psi_0}$ takes the following form
\begin{equation}
	\ket{\Psi_0} \simeq e^{T}\ket{\Psi_\mathrm{HF}},
	\quad
	\text{where}
	\quad
	T = \sum_{n} T_n
	\quad
	\text{and} 
	\quad 
	T_n = \frac{1}{(n!)^2} \sum_{\substack{i_1,\ldots,i_n \\ a_1,\ldots,a_n}} t^{i_1,\ldots,i_n}_{a_1,\ldots,a_n} a_{a_1}^{\dagger}{\ldots}a_{a_n}^{\dagger}a_{i_n}{\ldots}a_{i_1}
	\label{eq:grcc}
\end{equation}
for some real coefficients $t^{\cdots}_{\cdots}$, referred to as cluster amplitudes. 
Similarly, the $K$-th excited equation-of-motion (EOM) CC trial wavefunction $\ket{\Psi_K}$ is approximated by applying a linear excitation operator $R_{K}$ to the ground-state CC wavefunction
\begin{equation}
	\ket{\Psi_K} \simeq R_K\ket{\Psi_0} = R_K e^{T}\ket{\Psi_\mathrm{HF}},
	\quad
	\text{where}
	\quad
	R_{K} = \sum_{n} R_{K,n}
	\quad
	\text{and}
	\quad
	R_{K,n} = \frac{1}{(n!)^2} \sum_{\substack{i_1,\ldots,i_n \\ a_1,\ldots,a_n}} r^{i_1,\ldots,i_n}_{a_1,\ldots,a_n}(K) a_{a_1}^{\dagger}{\ldots}a_{a_n}^{\dagger}a_{i_n}{\ldots}a_{i_1}
	\label{eq:excc}
\end{equation}
for some real coefficients $r^{\cdots}_{\cdots}(K)$, referred to as excitation amplitudes.
Truncating $T$ and $R_{K}$ expansions at values of $n < \eta$ leads to the hierarchy of CC and EOMCC approximations.
Currently, the CC model with singles and doubles (CCSD, i.e., $T=T_1 + T_2$)~\cite{purvis82_1910} and the EOMCC formalism with singles and doubles (EOMCCSD, i.e., $T=T_1 + T_2$ and $R_{K}= R_{K,1} + R_{K,2}$)~\cite{bartlett89_57,bartlett93_414,stanton93_7029} methods in NWChem can be used to generate an initial wavefunction.
The trial wavefunction outputs are simple and reasonable approximations to~\autoref{eq:grcc} and~\autoref{eq:excc} which takes the following forms:
\begin{equation}
	\ket{\Psi_0} \simeq (1+T_1+T_2)\ket{\Psi_\mathrm{HF}}
	\label{eq:grini}
\end{equation}
in the case of the ground state and
\begin{equation}
	\ket{\Psi_K} \simeq (R_{K,1}+R_{K,2})\ket{\Psi_\mathrm{HF}}
	\label{eq:exini}
\end{equation}
for excited states.
Importantly, the cluster and excitation amplitudes in~\autoref{eq:grini} and~\autoref{eq:exini} are obtained from full CC/EOMCC calculations.
For more strongly correlated cases, one can envision the inclusion of higher rank excitations and products of cluster and/or excitation operators.
A typical input for generating one- and two-electron integrals---as well as leading CCSD and EOMCCSD trial wavefunction amplitudes---uses the NWChem TCE module and takes the form in~\autoref{lst:NWChem_Broombridge}. 
\begin{lstlisting}[
style=nwchem,label={lst:NWChem_Broombridge},caption={NWChem input deck for generating trial coupled-cluster wavefunctions.\hfill\texttt{examples/lih.nw}},
linerange={48-62}
]
# Tensor contraction engine module. As not all features of NWChem are currently supported by the Quantum Development Kit chemistry library, users are advised to use settings identical to the following, except for the //*{\color{comment-color}\texttt{nroots}}*// parameter.
tce
# Specification of the coupled-cluster model (CCSD)
ccsd
# Molecular orbital tile size defines the size of the blocks containing subsets of spin-orbital indices of the same spin and spatial symmetries. Currently the interface between NWChem and the Quantum Development Kit uses size 1, which is also its default value.
tilesize 1
# The two- and four-index transformations defined by the following keywords are performed at the orbital level assuming the same orbitals for both spin-up and spin-down spin-orbitals. This enables one to use the restricted Hartree--Fock and restricted open-shell Hartree--Fock references. The current version does not offer the unrestricted Hartree--Fock type one- and two-electron integrals.
2eorb
2emet 13
# Number of excited states to be calculated in the EOMCC model. If the this parameter is skipped, only the ground-state coupled-cluster calculations will be performed.
nroots 5
# Convergence threshold for CCSD and EOMCCSD calculations
thresh 1.0e-6
# End of tensor contraction engine module input group.
end
\end{lstlisting}

The trial wavefunctions~\autoref{eq:grini} and~\autoref{eq:exini} information along with the corresponding one- and two-electron integrals are printed out after a final specification of parameters outlined in the following~\autoref{lst:NWChem_Broombridge_imporant}.
\begin{lstlisting}[
style=nwchem,label={lst:NWChem_Broombridge_imporant},caption={NWChem input deck for additional outputs for Broombridge serialization.\hfill\texttt{examples/lih.nw}},
linerange={64-74}
]
# Instruct NWChem to print one- and two-electron integrals to the output file. This parameter also triggers remaining elements of calculations needed to characterize wavefunctions at the CCSD/EOMCCSD level.
set tce:print_integrals T
# Define the total number of correlated active orbitals in the calculation.
set tce:qorb 6
# Number of correlated active alpha (spin-up) electrons.
set tce:qela 2
# Number of correlated active beta (spin-down) electrons.
set tce:qelb 2

# NWChem task definition (compute coupled-cluster using TCE module)
task tce energy
\end{lstlisting}

%% file: review_simulation.tex
Quantum simulation is the original and perhaps most promising application of quantum computation~\cite{feynman2018feynman}. 
Given a Hamiltonian $H$ describing the system of interest and an evolution time $t$, the goal is to output the time-evolution operation $e^{-iHt}$. 
This time-evolution operator describes quantum dynamics as it evolves a the quantum state $\ket{\psi(0)}$ to the state $\ket{\psi(t)}=e^{-iHt}\ket{\psi(t)}$ at some future time $t$ in accordance to the time-dependent Schr\"odinger equation
\begin{align}
    i\frac{\dd}{\dd t}\ket{\psi(t)}=H\ket{\psi(t)}.
\end{align}
Importantly, $e^{-iHt}$ must be expressed in terms of quantum gates that may be implemented on a universal quantum computer. 
As this is difficult to do exactly, a maximum simulation error $\epsilon$ is allowed and we instead output an operation $U$ such that
\begin{equation}
    \|e^{-i H t} -U\| \le \epsilon.
    \label{eq:simprob}
\end{equation}
This criterion suffices to guarantee that the error in simulation for any initial quantum state is at most $\epsilon$.
Similarly, it also guarantees that the error in eigenvalues of $U$ are at most $\epsilon$~\cite[Theorem VI.3.11]{Bhatia1997MatrixAnalysis} from the exact time-evolution operator $e^{-iHt}$, which in turn encodes the eigenvalues of $H$.
The spectral norm is one of the most common choices for $\| \cdot \|$, although other choices are possible.

The complexity of simulating a second-quantized Hamiltonian on a classical computer scales exponentially with the number of spin-orbitals in the thermodynamic limit.
However, if we use a quantum computer, the dynamical simulation problem in~\eqref{eq:simprob} can be solved using a polynomial number of quantum operations.
This involves mapping the fermionic operators of the original Hamiltonian in~\autoref{eq:hsecond} to a Hamiltonian expressed by qubit operators, often by a Jordan--Wigner transformation, followed by an explicit algorithm for synthesizing $U$.
At present, the two most popular methods for constructing the operator $U$ are Trotter--Suzuki methods~\cite{Berry2007Efficient} and Qubitization~\cite{Low2016Qubitization}.
The former requires fewer qubits and can in practice require fewer gates for certain simulation problems, whereas the latter has better asymptotic scaling.
The precise number of gates needed for these simulations (as a function of the number of spin-orbitals) have fallen precipitously.
Early work suggested that the number of gate operations should scale like $O(N^{11})$~\cite{Wecker2014}, where $N$ is the number of spin-orbitals in the problem.
More recent work has reduced this to at most $O(N^6)$~\cite{Babbush2016Exponential,poulin2014trotter} for generic problems in chemistry or $O(N^2)$~\cite{Low2018IntPicSim} or lower for Coulomb interactions in the more structured plane-wave basis.
The time-evolution operator of dynamical simulation is usually used as a primitive in other algorithms.
For instance, static properties of a quantum system, such as the ground-state energy of a molecule, can be extracted using a phase estimation algorithm on $U$.
This returns an eigenphase $\tilde E_j t$ of $U$, up to some specified number of bits of precision in the algorithm.
This eigenphase of $U\ket{\Psi_j}=e^{i \tilde E_j t}\ket{\Psi_j}$ is selected with a probability $\Pr{[j]}=|\braket{\Psi_j|\Psi}|^2$ that depends on the overlap between the desired eigenstate and the prepared trial wavefunction $\ket{\Psi}$.
This trial wavefunction $\ket{\Psi}$ may be prepared by various means, like using another unitary similar in structure to $U$ such as the recently developed downfolding technique based on the extension of the subsystem-embedding-subalgebras to unitary coupled-cluster formalisms outlined in~\autoref{sec:review_nwchem}.
%

%% file: review_jordan_wigner.tex
Before simulating electronic structure problems, it is necessary to map the fermionic operators into Pauli operators that properly respect fermion anti-commutation relations and may be performed on a qubit quantum computer.
A number of such mappings from second-quantization, such as Bravyi-Kitaev, are possible~\cite{Seeley2012Bravyi} but a careful analysis shows no advantage over the simplest:  the Jordan--Wigner transformation~\cite{Hastings2015chemistry}.
The encoding of the fermionic operators into Pauli operators works as described below.
We define the state $\ket{0}_k$ to be the vacuum state for spin-orbital $k$ and similarly $\ket{1}_k$ is the occupied state.
It is then easy to see from~\autoref{eq:pauli} that
\begin{equation}
    a^\dagger_0 \rightarrow ({\X}_0 -i{\Y}_0)/2,\qquad a_0 \rightarrow ({\X}_0 +i{\Y}_0)/2.
\end{equation}
The creation operators acting on spin-orbitals with labels greater than zero need to be slightly modified to make sure that they properly anti-commute.
This can be achieved by noting that $\X \Z = -\Z \X$ and $\Y \Z = - \Y \Z$; that is, $\Z$ anti-commutes with both Pauli operators in the Jordan--Wigner decomposition of $a^\dagger_0$.
This means that we can construct creation operators acting on spin orbitals with an index greater than $0$ by attaching strings of $\Z$ operators to each of the qubits of lower labels to generate the proper anti-commutation relationship.
Specifically, we replace
\begin{equation}
    a^\dagger_p \rightarrow \prod_{j=0}^{p-1} {\Z}_j \left({\X}_p - i {\Y}_p \right)/2.\label{eq:jw}
\end{equation}
The Jordan--Wigner transformation of $a_p$ can be found by taking the adjoint of~\eqref{eq:jw}.

As an example, consider the term $a^\dagger_j a_j a^\dagger_k a_k = (\id - {\Z}_j + {\Z}_k + {\Z}_j {\Z}_k)/4$ after the Jordan--Wigner transformation. 
Time-evolution by this term may be decomposed into the sequence of quantum gates
\begin{align}
    e^{-i a^\dagger_j a_j a^\dagger_k a_k t} &= e^{-it/4}e^{i{\Z}_jt/4}\cdot e^{i{\Z}_kt/4}\cdot e^{-i{\Z}_j{\Z}_k t/4}\nonumber\\
    &= e^{-it/4}e^{i{\Z}_jt/4}\cdot e^{i{\Z}_kt/4} \cdot \Cnot_{j,k} \cdot e^{-i{\Z}_k t/4} \cdot \Cnot_{j,k}.
\end{align}
Note that because each term commutes in the Jordan--Wigner representation of the Hamiltonian, this expression is exact.
In contrast, if $a^\dagger_j a_k + a^\dagger_k a_j = {\X}_j {\Z}_{j+1} \cdots {\Z}_{k-1} {\X}_k/4 + {\Y}_j {\Z}_{j+1} \cdots {\Z}_{k-1} {\Y}_k/4$, then time evolution by this term decomposes into 
\begin{align}
    e^{-i(a^\dagger_j a_k + a^\dagger_k a_j)t} &= \Had_j\cdot \Had_k\cdot \Cnot_{j,k} \cdots \Cnot_{k-1,k}\cdot e^{-i {\Z}_k t/4}\cdot\Cnot_{k-1,k}\cdots\Cnot_{j,k}\cdot \Had_j \cdot\Had_k\nonumber\\
    & \qquad \cdot \Had_j \cdot\Had_k \cdot \Phase_j^\dagger \cdot \Phase_k^\dagger \cdot \Cnot_{j,k} \cdots \Cnot_{k-1,k} \cdot e^{-i {\Z}_k t/4} \cdot\Cnot_{k-1,k}\cdots\Cnot_{j,k} \cdot \Phase_j \cdot \Phase_k
    \nonumber\\
    &\qquad\cdot \Had_j \cdot\Had_k.
\end{align}
As before, this expression is also exact as each term in the Pauli representation commutes.

These long sequences of controlled-\Not~ gates are needed to ensure the anti-commutation rules of fermionic creation operators and are collectively known as Jordan--Wigner strings.
Specific forms for the evolution operators corresponding to each of the different one- and two-electron operators in the second-quantized Hamiltonian are well known~\cite{whitfield2011simulation} and implementations of these gates are provided within the Hamiltonian simulation library.

%% file: review_trotter.tex
In essentially all cases of interest, the terms of a Hamiltonian $H$ representing an electronic structure problem do not commute in either the fermion or Pauli representations. 
Trotter--Suzuki methods, often abbreviated to just ``Trotter'', are a cornerstone method for simulating non-commuting Hamiltonians on quantum computers. 

In general, it is difficult to compile $e^{-iHt}$ into quantum gates directly; however, $H$ is often the sum of a large number of individual terms $H_j$ such that it is easy to find circuits for each $e^{-iH_j t}$.
In the simplest example, if $H= a\X + b\Y+ c \Z$, then each of these individual terms can be simulated using $R_\X$, $R_\Y$, and $R_\Z$ gates (which are defined in~\autoref{sec:reviewQC}).
In general, if a Hamiltonian is of the form $H=\sum_{j=1}^M h_j P_j$ where each $P_j$ is a tensor product of Pauli operators Hermitian, then a Trotter--Suzuki approximation can be written as
\begin{equation}
    \label{eq:trotter_first_order}
    e^{-iHt} = \prod_{j=1}^M e^{-ih_j P_j t} + \mathcal{O}(M^2  \max_{j,k}\|[P_j,P_k]\| \max|h_j|^2t^2).
\end{equation}
One can see that the approximation error is controlled by the Trotter step-size $t$, which should suitably small to reach chemical accuracy.
Quite often, the term Trotter number $1/t$ is also used, which is the number of Trotter--Suzuki formula applications required to achieve unit-time time-evolution.
Furthermore, elementary quantum circuits involving chains only Clifford gates and a single qubit rotation can be used to simulate each exponential of a Pauli operator.
Therefore, if the chemical Hamiltonian can be decomposed into a sum of a modest number of quantum circuit is known to simulate $e^{-iHt}$, then the Trotter formula
can be used to build an approximation to $e^{-iHt}$ assuming $t$ is sufficiently small.
Using mappings such as Jordan--Wigner, fermion Hamiltonians are represented by as a sum of Pauli operators, and methods exist for simulating such exponentials using a polynomial number of primitive quantum gates.

One example of a non-commuting fermion Hamiltonian is a combination of the terms from the previous section.
Let $H = a^\dagger_j a_j + a^\dagger_j a_k + a^\dagger_k a_j = (\id - {\Z}_j)/2 + {\X}_j {\Z}_{j+1} \cdots {\Z}_{k-1} {\X}_k/4 + {\Y}_j {\Z}_{j+1} \cdots {\Z}_{k-1} {\Y}_k/4$.
A simulation circuit can be formulated using exactly the same methodology.
However, because $[{\Z}_k,{\X}_k]\ne 0 \ne [{\Z}_k,{\Y}_k]$ the error---often called the Trotter error---is $O(t^2)$ for such a simulation.
Specifically, it can be shown using the same approach demonstrated above that
\begin{align}
    e^{-iHt} &= e^{-it/2}e^{-i{\Z}_jt/2}\cdot \Had_j\cdot \Had_k\cdot \Cnot_{j,k} \cdots \Cnot_{k-1,k}\cdot e^{-i {\Z}_k t/4}\cdot\Cnot_{k-1,k}\cdots\Cnot_{j,k}\cdot \Had_j \cdot\Had_k\nonumber\\
    & \qquad \cdot \Had_j \cdot\Had_k \cdot \Phase_j^\dagger \cdot \Phase_k^\dagger \cdot \Cnot_{j,k} \cdots \Cnot_{k-1,k} \cdot e^{-i {\Z}_k t/4} \cdot\Cnot_{k-1,k}\cdots\Cnot_{j,k} \cdot \Phase_j \cdot \Phase_k \cdot \Had_j \cdot\Had_k
    \nonumber\\ 
    &\qquad+ O(M^2 \max_{j,k}\|[H_j,H_k]\|t^2),
\end{align}

Higher-order Trotter--Suzuki decompositions~\cite{Berry2007Efficient} also exist and are available within the Hamiltonian simulation library.
The simplest such decomposition is the symmetric (or second order) Trotter formula, which takes the form
\begin{equation}
    e^{-iHt} = \prod_{j=1}^M e^{-iH_j t/2} \prod_{j=M}^1 e^{-iH_j t/2} + O(M^3 \max_{j,k,\ell}\|H_{\ell}\|\|[H_j,H_k]\|t^3).
\end{equation}
Arbitrarily high-order Trotter formulas can also be constructed from the second-order formula; however, the highest-order formula used in practice is the fourth-order formula~\cite{Childs2017Speedup}.

Once $e^{-iHt}$ has been decomposed into a product of elementary unitary operations using one of the above formulas, we have everything that we need to simulate quantum dynamics on a quantum computer. 
However, for chemistry simulation applications, we are usually interested in static properties like the correlation energy of a molecule.
Observables such as eigenvalues of the Hamiltonian can then be extracted from the time-evolution operator.  
For example, if $H\ket{\psi}= E\ket{\psi}$ then $e^{-iHt}$ has eigenvalue $e^{-iEt}$ on the state.
Thus, if we apply phase estimation on $e^{-iHt}$, then we can learn $Et$ directly from the phase and in turn $E$ since $t$ is known.
For the purpose of estimating the ground-state eigenvalue, the second-order Trotter formula yields the same accuracy as the first-order Trotter formula, which allows a potential savings of a factor of $2$ in the complexity~\cite{poulin2014trotter}.
However, most simulation results use the symmetric formula to simplify the error analysis.

In theory, this simulation approach does not compare favorably asymptotically to methods based on qubitization or linear combinations of unitaries; however, Trotter--Suzuki formulas require fewer qubits than any other known method.
Moreover, the complexity of the simulation depends strongly on the size of the commutators between the Hamiltonian terms.
In practice, this means that Trotter--Suzuki methods can be more efficient than more recent simulation methods for some problems~\cite{Childs2017Speedup} and, therefore, will remain an important part of the landscape of quantum simulation algorithms for the foreseeable future.

%% file: review_qubitization.tex
In the previous section, we described the Trotter--Suzuki algorithm that directly approximates the unitary time-evolution operator $e^{-iHt}$.
However, for the purposes of estimating eigenvalues of the Hamiltonian $H$, it suffices to implement time-evolution by any monotonic function of $H$, say $e^{if(H)}$, where $f(\cdot)$ is applied to the eigenvalues of $H$ without modifying the eigenvectors~\cite{Poulin2018Spectral,babbush2017sorting}.
Qubitization~\cite{Low2016Qubitization} is a simulation technique for synthesizing a unitary that is exactly $e^{i\sin^{-1}(H/h)}$ for some normalization constant $h\ge\|H\|$, up to fixed phase factors and local isometries.
Compared to the Trotter--Suzuki algorithm, qubitization offers a different complexity tradeoff that may be advantageous in certain situations.

The starting point of qubitization is a Hamiltonian represented as a linear combination of $N$ Pauli operators $P_j$ with positive coefficients $h_j$, say
\begin{align}
    H = \sum_{j=0}^{N}h_j P_j.
\end{align}
Such a decomposition can be found for the chemistry Hamiltonian by using the Jordan--Wigner decomposition given in~\autoref{eq:jw}.
Information about this Hamiltonian is then encoded in two unitary operators $\Prepare$ and $\Select$. 
Coefficient information is encoded in a quantum state 
\begin{align}
    \ket{h}=\sum_{j=0}^{N-1}\sqrt{\frac{|h_j|}{h}} \ket{j},
\end{align}
where $h=\sum_{j=0}^{N-1}|h_j|$ is the one-norm of coefficients. 
Note that the number state $\ket{j}$ encodes a binary representation of $j$, e.g. $\ket{6}=\ket{1}\ket{1}\ket{0}$.
This state can be prepared by the quantum circuit
\begin{align}
    \label{eq:prepare_sbm}
    \Prepare \ket{0} = \ket{h},
\end{align}
implemented following Ref.~\cite{shende2006synthesis}. 
Operator information is encoded in a unitary operator, implemented following Ref.~\cite{Babbush2018encoding}, 
\begin{align}
    \label{eq:select_default}
    \Select = \sum_{j=0}^{N-1}\ket{j}\bra{j} \otimes P_j,
\end{align}
that applies the $j^{\textrm{th}}$ Pauli operator given the number state $\ket{j}$.
Together, these combine to apply the Hamiltonian in the sense of
\begin{align}
    \label{eq:block_encoding}
    V&=(\Prepare^\dag\otimes \I)\cdot\Select\cdot(\Prepare\otimes \I)\\\nonumber
    \frac{H}{h}    &=(\bra{0}\otimes\I) \cdot V \cdot (\ket{0}\otimes\I).
\end{align}

By combining $\Prepare$ and $\Select$, qubitization is simply the walk operator
\begin{align}
    W=((2\ket{0}\bra{0}-\I)\otimes\I)\cdot V.
\end{align}
When evaluating its action on eigenstates $\ket{E}$ of $H\ket{E}=E\ket{E}$ with energy $E$ when the input in the other register is $\ket{0}$, there are two cases of interest: $|E|=h$, and $|E|\neq h$. 
In the former case,
\begin{align}
    W\ket{0}\ket{E}=\frac{E}{h}\ket{0}\ket{E}=\operatorname{sign}[E]\ket{0}\ket{E}.
\end{align}
Thus, the walk operator simply applies a phase $0$ or $\pi$ to the input. 
In the latter case, using the simplified notation $\frac{E}{h}=\lambda$, 
\begin{align}
    W\ket{0}\ket{E}&=\lambda\ket{0}\ket{E}-\sqrt{1-|\lambda|^2}\ket{0E^\perp},
\end{align}
where $\ket{0E^\perp}$ is orthogonal to the original input. 
By rearranging, this new state
\begin{align}
    \ket{0E^\perp} = \frac{- W\ket{0}\ket{E} + \lambda\ket{0}\ket{E} }{\sqrt{1-|\lambda|^2}}.
\end{align}
Thus, we may evaluate the matrix elements of $W$ in this basis as
\begin{align}
    W=\bigoplus_E\begin{bmatrix}
    E/h& \sqrt{1-|E/h^2}\\
    -\sqrt{1-|E/h^2} & E/h
    \end{bmatrix}.
\end{align}
By diagonalizing each subspace separately, we see that $W$ applies a phase $ e^{\mp i\cos^{-1}{(E/h)}}$ to the eigenstates $\frac{\ket{0}\ket{E}\pm\ket{0E^\perp}}{\sqrt{2}}$.
Within this basis, the spectrum of the walk operator is isomorphic to
\begin{align}
    W=e^{-i\Y\otimes \cos^{-1}(H/h)}=-i e^{-i\Y \otimes\sin^{-1}(H/h)}.
\end{align}

%% file: review_qubitization_chem_optimized.tex
In the previous section, we described the qubitization algorithm and presented a generic implementation.
Here, we state some basic circuit optimizations we have implemented that reduce the overall complexity.
First, we consider state-preparation optimizations that reduce the $\T$-gate complexity, but at the expense of introducing additional qubits.
Second, we consider optimizations that exploit the structure of fermion Hamiltonians.

The default state preparation circuit $\Prepare$ in~\autoref{eq:prepare_sbm} for $\ket{\alpha}$ is implemented following the procedure by Shende, Bullock, and Markov~\cite{shende2006synthesis}.
For a quantum state of dimension $N$, this requires $\lceil\log_2{N}\rceil$ qubits and $\mathcal{O}(N)$ arbitrary $Z$ rotations.
By synthesizing each rotation with $\T$ gates~\cite{Ross2016Optimal}, the overall $\T$ gate complexity for approximating $\ket{\alpha}$ to error $\epsilon$ is $\mathcal{O}(N \log{(N/\epsilon)})$. 

A challenge faced by this state-preparation method is that the cost of the algorithm scales multiplicatively with $\log(1/\epsilon)$.
If an exacting estimate of the error is required, then the number of rotation gates needed can be prohibitively large if the $L_1$ norm of the Hamiltonian terms or the error tolerance is small.  
This situation can be ameliorated by including ancillary qubits, the use of which allows us to have error scaling that is additive (rather than multiplicative) in the desired error tolerance.
Specifically, we aim to synthesize a state $\ket{h'}$ such that
\begin{align}
    \label{eq:qrom}
    \ket{h'}=\sum_{j=0}^{N-1}\sqrt{\frac{|h_j|}{h}}\ket{j}\ket{\mathrm{garb}_j},
\end{align}
where $\ket{\mathrm{garb}_j}$ is some additional arbitrary ``garbage state'' entangled with $\ket{j}$ and $h$ is the $L_1$ norm of the coefficient vector with entries $h_j$.
One may verify that, even with this garbage state, the defining equation~\autoref{eq:block_encoding} for block-encoding the Hamiltonian returns an identical result.
This state may be synthesized using the QROM procedure~\cite{Babbush2018encoding,Low2018Trading}, at cost $\mathcal{O}(n\log{(1/\epsilon)})$ qubits, and $\mathcal{O}(N+\log{(N/\epsilon)})$ $\T$ gates. 

The default $\Select$ circuit in~\autoref{eq:select_default} applies an arbitrary Pauli operator conditioned on one of $N$ different control inputs, and has a $\T$-gate cost of less than $4N$~\cite{Babbush2018encoding}. 
In a fermionic Hamiltonian with $M$ spin-orbitals, there may be up to $N=\mathcal{O}(M^4)$ different Pauli operators, corresponding to a spin representation of the two-body terms.
However, these Pauli operators in fermion Hamiltonians are highly structured. 
For instance, the Jordan--Wigner representation in~\autoref{eq:jw} expresses the fermion operator $a_p$ as the sum of $X$ and $Y$ Pauli on qubit $p$ followed by a string of $Z$ Paulis from qubits $0$ to $p-1$. 
The circuit selecting these structured Pauli operators can be synthesized using $\mathcal{O}(M)$ $\T$ gates, following the method of~\cite{Babbush2018encoding}.  


%% file: algo.tex

An important primitive in quantum simulation applications is obtaining an eigenvalue of some given unitary operator $U\ket{\psi}=e^{i\theta}\ket{\psi}$ with eigenstate $\ket{\psi}$.
As the eigenvalues are always of the form $e^{i\theta}$ for some phase $\theta$, the process has earned the moniker ``phase estimation''.
The origins of phase estimation actually predate quantum mechanics and the quantum algorithm by many years.
The first such example of phase estimation is the Mach--Zehnder interferometer \cite{Zetie2000MachZehnder}, wherein a phase delay is put in one of two arms of an interferometer.
At the end of the protocol, light heading down both paths is allowed to interfere and the phase difference between the two paths that the light travels through becomes immediately obvious from the interference pattern.  
As we will shortly find out, quantum algorithms for phase estimation are conceptually similar. 

One large family of algorithms are based on a procedure called iterative phase estimation. 
This procedure requires a qubit that stores the two paths in the analogous interferometer and a controlled quantum circuit controlled-$U$ such that
\begin{align}
    \text{controlled-}U\ket{0}\ket{\psi} = \ket{0} \ket{\psi},
    \quad
    \text{and}
    \quad
    \text{controlled-}U\ket{1}\ket{\psi} = \ket{1} U\ket{\psi}= e^{i\theta} \ket{1}\ket{\psi}.
\end{align}
In other words, the controlled unitary only applies its phase to the portion of the quantum state in the ``$1$'' branch.  
Iterative phase estimation splits the quantum state uniformly over both branches through a Hadamard gate, applies the controlled unitary, and then recombines the quantum state by applying a Hadamard again to the path qubit as follows.
\begin{align}
    \ket{0}\ket{\psi} &\mapsto \frac{1}{\sqrt{2}} \left( \ket{0} \ket{\psi}  + \ket{1}\ket{\psi} \right)\nonumber\\
    &\mapsto \frac{1}{\sqrt{2}} \left( \ket{0} \ket{\psi}  + e^{i\theta}\ket{1}\ket{\psi} \right)\nonumber\\
    &\mapsto \frac{1}{2} \left( (1+e^{i\theta})\ket{0} \ket{\psi}  + (1-e^{i\theta})\ket{1}\ket{\psi} \right)\nonumber\\
    &=e^{i\theta/2} \left( \cos(\theta/2)\ket{0} \ket{\psi}  -i \sin(\theta/2)\ket{1}\ket{\psi} \right)
    \label{eq:singlepass}
\end{align}
Thus the phase can be inferred by measuring the path qubit many times, estimating the probability $\Pr{[0]}=\cos^2{(\theta/2)}$ and computing $\theta \approx 2\cos^{-1}(\sqrt{\Pr{[0]}})$.  
Of course, this is terribly inefficient.

In practice, phase estimation is almost always implemented as a multi-pass algorithm, meaning that controlled-$U$ is applied multiple $L$ times before each measurement.
Furthermore, the process of inferring the feedback phase $\phi$ is made easier by also rotating the path qubit at each iteration; although this does not provide additional information about $\theta$, using a feedback phase $\phi$ can help with the stability of numerical implementations.
Combining both of these together yields the transformation
\begin{equation}
    \ket{0}\ket{\psi} \mapsto \left( \cos(L(\theta+\phi)/2)\ket{0} \ket{\psi}  +i \sin(L(\theta+\phi)/2)\ket{1}\ket{\psi} \right),
    \label{eq:multipass}
\end{equation}
up to an irrelevant global phase. 
We may quantify the effectiveness of this procedure by the Fisher information $I$, whose inverse which lower bounds the variance of any unbiased estimator of $\theta$. 
A straightforward calculation shows that the Fisher information $I$ scales quadratically with the evolution time, $L^2$~\autoref{eq:multipass}, whereas $I$ scales linearly with the number of times each measurement is repeated.
Informally, this is the distinction between the ``Heisenberg'' and ``standard quantum'' limits, respectively~\cite{Ferrie2013Periodic}.

The art in phase estimation is to choose a policy for picking a sequence of experiments with varying $L$ and $\phi$. 
The measurement records can be combined on a classical computer to yield a useful estimate of the phase, whose distribution depends on this choice.
Typically, the phase estimate is obtained by a maximum likelihood estimator.
However, other estimators such as the posterior mean can be applied in Bayesian approaches that exploit prior information on the distribution of $\theta$.
Regardless, it is known that a judicious sequence of $L$ and $\phi$ allows the uncertainty in the estimated $\theta$ to scale as $\Delta\propto 1/L_{\exp}$, where $L_{\exp}=\sum_{j}L_j$ is the total number of times the controlled unitary is applied across all experiments.
This yields a quadratic advantage relative to na\"{\i}ve statistical sampling, wherein the uncertainty scales as $1/\sqrt{L_{\exp}}$.  
The quadratic advantage provided by the phase estimation approach is optimal; otherwise, the Heisenberg uncertainty principle for photon number and phase could be violated.
After fixing an eigenstate of the unitary to be queried by a phase estimation algorithm, it is known that the Cramer--Rao bound is tight for efficient phase estimators.
Thus the number of queries required to obtain a target error $\Delta$ in standard deviation in the asymptotic limit is simply
\begin{align}
    \label{eq:pe_cramer-rao}
    \text{Number of queries}=L_{\exp}\approx\frac{1}{\Delta}.
\end{align}

The analysis of phase estimation given above comes with a major caveat: the input state $\ket{\psi}$ is assumed to be an eigenstate of $U$, which is almost never true.  
In this situation, phase estimation can be viewed as performing a measurement in the eigenbasis of the unitary $U$.
This means that if $\ket{\psi} = a \ket{\psi_0} + b \ket{\psi_1}$ for eigenstates $\ket{\psi_0}$ and $\ket{\psi_1}$ then the probability of measuring the eigenvalue corresponding to $\ket{\psi_0}$ is $|a|^2$.
Learning a particular eigenphase with high probability therefore also requires preparing an initial state that has a sufficiently large overlap with the target state.
An important mitigation strategy for this phenomena is to disregard output from any phase estimation that yields a result that disagrees with prior estimates for the eigenvalue.
This strategy is enabled within our schema, described next in~\autoref{sec:representation}, by allowing users to specify upper and lower bounds on eigenvalue of interest.

Note that in this approach to phase estimation, the number of times that $U$ is applied is always an integer.
Choosing an integer number of applications means that eigenphase $\theta$ and $\theta + 2m\pi$, for any integer $m$, yields precisely the same likelihood and thus cannot be distinguished.
This effect is called ``wrap-around.''
Wrap-around can be dealt with by using methods such as various flavors of Bayesian phase estimation that use a non-integer number of queries to the unitary, which can be realized using fractional query techniques~\cite{Gilyen2018singular}.
In practice, this issue seldom occurs in phase estimation of time-evolution operators as either the total phase needs to be kept small to control errors or the maximum phase that can be observed is $\pi/2$, depending on the simulation method of choice.
For this reason, and the fact that it comes with provable bounds on the uncertainty and failure probability, we focus on a form of phase estimation called robust phase estimation~\cite{Kimmel2015Robust} that only uses an integer number of queries to controlled-$U$. 
Other algorithms exist and we recommend the interested reader to look at faster phase estimation~\cite{Svore2014fasterphase}, Bayesian phase estimation~\cite{wg_efficient_2015}, and quantum phase estimation~\cite{Nielsen2004}.
Each of these approaches has different tradeoffs between experimental run time, classical processing, and the number of quantum bits used in the protocol.

%% file: representation.tex

The Broombridge schema\footnote{So named after the Broom bridge in Dublin, Ireland, upon which Sir William Rowan Hamilton, the namesake of `Hamiltonian', inscribed the first defining equations of quarternions.} defines a data structure for representing electronic structure problems together with supporting metadata to enable effective simulation on a quantum computer.
Using a human-readable serialization, this provides an interface between electronic structure calculation tools, in particular NWChem, and the Microsoft Quantum Development Kit chemistry library.
By standardizing this interface under the open-source MIT license, we also enable potential inter-operation between any set of classical and quantum chemistry simulation software packages, and enable future schema extensions to meet the requirements of state-of-the-art electronic structure algorithms.

We outline in~\autoref{sec:broombridge_components} the essential components contained in Broombridge that are relevant to the chemistry library.
Subsequently, we describe in~\autoref{sec:broombridge_generation} how Broombridge may be generated by NWChem, which is used later in the examples of~\autoref{sec:applications}.

\subsection{Broombridge v0.1 specifications}
\label{sec:broombridge_components}

We now present snippets from the Broombridge example of LiH that highlight its essential keys and values.
As future versions of Broombridge may not be backwards-compatible, each Broombridge instance begins with a version number and a link to its specification.
Some entries of Broombridge are required and will not pass validation if omitted, whereas other entries are optional metadata, as shown in the following~\autoref{lst:broombridge_format}.
\begin{lstlisting}[language=yaml,label={lst:broombridge_format},caption={Broombridge version number formatting.\hfill\texttt{examples/lih.yaml}},
linerange={1-9},
]
# (Required) Address to complete definition for validaing Broombridge v0.1. 
"$schema": https://raw.githubusercontent.com/Microsoft/Quantum/master/Chemistry/Schema/broombridge-0.1.schema.json

# (Required) Broombridge version of electronic structure problem representation. 
format: {version: '0.1'}

# (Optional) Bibliography for problems represented in this Broombridge. 
bibliography:
- url: 'https://www.nwchem-sw.org'
\end{lstlisting}

A quantitative description of the electronic structure problem is stored as an entry in the `integral\_set' list -- multiple Broombridge problems may be stored in this list.
Each entry in this list contains a description of the problem.
Some parameters are essential for specifying a complete quantum simulation problem.
This includes the number of orbitals and electrons required, the constant energy offsets equivalent to identify terms in the Hamiltonian, the Hartree--Fock energy, and the one-electron and two-electron integrals over the defined orbital subspace.
These are outlined in the following~\autoref{lst:broombridge_integral_sets_required}.
\begin{lstlisting}[language=yaml,label={lst:broombridge_integral_sets_required},caption={Essential Broombridge components for specifying an electronic structure problem.\hfill\texttt{examples/lih.yaml}},
linerange={31-43,54-65}
]
  # Terms in the electronic structure Hamiltonian.
  hamiltonian:
  # List of one-electron integrals
    one_electron_integrals:
      # Only non-zero terms are specified in a sparse format.
      format: sparse
      units: hartree
      values:
      # Each element [i, j, //*{\color{comment-color}$h_{ij}$}*//], where //*{\color{comment-color}$i,j\in\{1,\cdots,\text{n\_orbitals}\}$}*//, is the orbital integral //*\autoref{eq:h_pq}*//. 
      # As //*{\color{comment-color}$h_{ij}=h_{ji}$}*//, only one of these terms must be specified.
      - [1, 1, -4.7225445389]
      - [2, 1, 0.1046509472]
      # etc.
   
    # List of two-electron integrals        
    two_electron_integrals:
      format: sparse
      index_convention: mulliken
      units: hartree
      values:
      # Each element [i, l, j, k, //*{\color{comment-color}$h_{ijkl}$}*//], where //*{\color{comment-color}$i,j,j,k\in\{1,\cdots,\text{n\_orbitals}\}$}*//, is the orbital integral //*\autoref{eq:h_pqrs}*//. 
      # As //*{\color{comment-color}$h_{ijkl}=h_{jilk}=h_{ikjl}=h_{lkji}$}*// for real orbitals, only one of these terms must be specified.
      - [1, 1, 1, 1, 1.6586341297]
      - [2, 1, 1, 1, -0.110636355]
      # etc.    
\end{lstlisting}

Some elements of this description are optional but are highly recommended as they can be used by the chemistry library.
Importantly, as shown in~\autoref{lst:broombridge_integral_sets_recommended}, this includes multi-configurational trial wavefunctions that are superpositions of Slater determinants, as well as approximations to the full configuration-interaction energy that can also be used to determine the principle range of phase estimation, if applicable.
\begin{lstlisting}[language=yaml,label={lst:broombridge_integral_sets_recommended},caption={Recommended Broombridge properties that can be used by the chemistry library.\hfill\texttt{examples/lih.yaml}},
linerange={164-193}
]
  # Approximate upper and lower bounds on the exact ground-state energy.
  fci_energy: {units: hartree, lower: -7.981844675840114, upper: -7.781844675840115}

  # Various multi-configurational trial wavefunctions.
  initial_state_suggestions:

  # This is a ground state trial wavefunction.
  - state:
      # Approximate energy of this state
      energy: {units: hartree, value: -7.881844675840115}
    
      # Label used to reference this state. 
      label: '|G>'
      
       # The trial wavefunction described by a superposition where each entry is the amplitude followed
       # by the corresponding Slater determininant described by fermion operators acting on the vacuum 
       # state. In Polish notation, (3a)+//*{\color{comment-color}$\equiv a^{\dagger}_{3,\uparrow}$}*//, (2b)//*{\color{comment-color}$\equiv a_{2,\downarrow}$}*//, (2a)//*{\color{comment-color}$\equiv a_{2,\uparrow}$}*// and so on.
      superposition:
      - [0.993182561322233, (1a)+, (2a)+, (1b)+, (2b)+, '|vacuum>']
      - [-0.116569292206005, (6a)+, (6b)+, (2b), (2a), (1a)+, (2a)+, (1b)+, (2b)+, '|vacuum>']

  # This is a first excited state trial wavefunction.
  - state:
      energy: {units: hartree, value: -7.750219591632256}
      label: '|E1>'
      superposition:
      - [0.664745837968804, (3a)+, (2a), (1a)+, (2a)+, (1b)+, (2b)+, '|vacuum>']
      - [0.664745837968804, (3b)+, (2b), (1a)+, (2a)+, (1b)+, (2b)+, '|vacuum>']
      - [-0.246552630006247, (3a)+, (3b)+, (2b), (2a), (1a)+, (2a)+, (1b)+, (2b)+, '|vacuum>']
      - [0.166489853209298, (3a)+, (6b)+, (2b), (2a), (1a)+, (2a)+, (1b)+, (2b)+, '|vacuum>']
\end{lstlisting}

Some elements of this description are optional and are not currently used by the chemistry library.
For instance, this includes the molecule geometry and the basis set as outlined in the following~\autoref{lst:broombridge_integral_sets_optional}.
Full details of all other fields are documented online at~\url{https://docs.microsoft.com/en-us/quantum/libraries/chemistry/schema/broombridge}. 
\begin{lstlisting}[language=yaml,label={lst:broombridge_integral_sets_optional},caption={Select optional Broombridge properties not used by the chemistry library.\hfill\texttt{examples/lih.yaml}},
linerange={257-270}
]
  # Basis set used to represent problem.
  basis_set: {name: sto-3g, type: gaussian}

  # Geometry describing the problem.
  geometry:
    atoms:
    - {name: Li, coords: [0.0, 0.0, -0.406]}
    - {name: H, coords: [0.0, 0.0, 1.218]}
    coordinate_system: cartesian
    symmetry: c1
    units: angstrom

  # Any additional miscellaneous metadata. This must be included, but may be empty. 
  metadata: {}
\end{lstlisting}

\subsection{Generating Broombridge with NWChem}
\label{sec:broombridge_generation}

One of the following methods may be used to either obtain or generate a description of an electronic structure problem, serialized as Broombridge.
\begin{itemize}
   \item The easiest way to obtain Broombridge is from the numerous existing samples at, say,~\url{https://github.com/Microsoft/Quantum/tree/master/Chemistry/IntegralData/YAML}.
   \item The next easiest way to generate Broombridge is to use the EMSL Arrows Builder for the Microsoft Quantum Development Kit at~\url{https://arrows.emsl.pnnl.gov/api/qsharp_chem}. 
   This is a web-based frontend to NWChem and chemical computational databases for many materials and chemical modeling problems via a broad spectrum of digital communications, including posts to web API.
   With this framework, a molecule can be input into EMSL Arrows using a variety of techniques documented at~\url{http://www.nwchem-sw.org/index.php/EMSL_Arrows} say, the simplified molecular-input line-entry system, a graphical 2D or 3D molecule builder, or as a standard NWChem input deck.
   \item For the most flexibility, PNNL also provides a Docker image that automatically compiles a virtual machine containing a complete and executable version of NWChem.
   \item An advanced user may also download and compile NWChem from source.
\end{itemize}
Broombridge is obtained by serializing the output file dump of an NWChem computation of the format outlined in~\autoref{sec:review_nwchem}, using the Python script provided with NWChem.
A convenient frontend to the Docker image for NWChem is provided with the Quantum Development Kit as a cross-platform PowerShell script.

%% file: simulation_introduction.tex
The Microsoft Quantum Development Kit chemistry library implements the quantum simulation algorithms of~\autoref{sec:review_of_quantum_simulation} in \qsharp~with chemistry-specific optimizations, and provides an interface in \csharp~to define fermion Hamiltonians relevant to chemistry, such as through the Broombridge schema in~\autoref{sec:representation}. 
Taken together, a quantum simulation of any electronic structure problem generated by NWChem in~\autoref{sec:broombridge_generation} may be executed on any the target machine provided by the Microsoft Quantum Development Kit. 
The two target machines relevant here are: (1) the full-state simulator, which emulates a noiseless quantum computer, albeit with exponential time scaling in qubit count, and (2) the trace simulator, which evaluates various resource costs of the simulation with polynomial time scaling in qubit count.

We describe use of this library through a quantum simulation of molecular hydrogen. 
In~\autoref{sec:simulation_construct_hamiltonian}, we construct the hydrogen Hamiltonian, and simulate its real-time evolution.
Real-time evolution is then invoked as a subroutine to obtain estimates of the ground-state energy in~\autoref{sec:simulation_estimate_energy}. 
For molecules with many qubits, full-state simulation on a classical machine is intractable. 
However, efficiently obtaining cost estimates of the simulation is possible simply by swapping in the trace simulator.

%% file: simulation_construct_hamiltonian.tex
\lstMakeShortInline[style=csharp]!

Consider a simple representation of molecular hydrogen in the $\textsc{sto-3g}$ with two orbitals.
In this basis, the hydrogen Hamiltonian has the form
\begin{align}
    \label{eq:Hydrogen_Hamiltonian}
    H=h_0I+\sum_{i,j}h_{i,j}\sum_{\sigma\in\{\uparrow,\downarrow\}} a^\dag_{i,\sigma}a_{j,\sigma}
    +\frac{1}{2}\sum_{i,j,k,l}h_{i,j,k,l}\sum_{\sigma,\rho\in\{\uparrow,\downarrow\}} a^\dag_{i,\sigma}a^\dag_{j,\rho}a_{k,\rho}a_{l,\sigma},
\end{align}
where the only non-zero entries are 
\begin{align}
    \label{eq:Hydrogen_coefficients}
    h_0&=0.71377618,\quad
    h_{0,0}=-1.252477495,\quad
    h_{1,1}= -0.475934275,\quad\\\nonumber
    h_{0,0,0,0}&= 0.674493166,\quad
    h_{0,1,0,1}= 0.181287518,\quad
    h_{0,1,1,0}= 0.663472101,\quad\text{and }\quad
    h_{1,1,1,1}= 0.697398010.
\end{align}
Note that the spin and orbital indices are written explicitly and we use zero-indexing for orbitals, which should be compared to the implicit notation of~\autoref{eq:hsecond} for the Hamiltonian, and~\autoref{eq:h_pq} and~\autoref{eq:h_pqrs} for the coefficients.
In total, there are two spin-orbitals occupied by two electrons.
It is also necessary to define the initial state acted on by the Hamiltonian.
In general, such states can be written as
\begin{align}
    \ket{\Psi}=\sum_{i_{1}<\cdots<i_{\eta}}\lambda_{i_{1},\cdots i_{\eta}}a^\dag_{i_{1}}\cdots a^\dag_{i_{\eta}}\ket{0},\quad \sum_{i_{1}<\cdots<i_{\eta}}|\lambda_{i_{1},\cdots i_{\eta}}|^2=1,
\end{align}
and following~\autoref{fciex}, are linear combinations of Slater determinants with a fixed number of $\eta$ electrons.
The Hartree--Fock state 
\begin{align}
    \ket{\Psi_\mathrm{HF}}=a^{\dag}_{0,\downarrow}a^{\dag}_{0,\uparrow}\ket{0}
\end{align}
is the simplest example, and for Hydrogen, approximates the true ground state reasonably well.

We start by importing the chemistry library in~\autoref{lst:import-csharp-library}.
\lstinputlisting[
style=CSharp,
gobble=0,
linerange={9-12},
label={lst:import-csharp-library},
caption={Importing the C\# component of the chemistry library.\hfill\texttt{examples/hydrogen/host.cs}}
]{\sourcefolder/examples/hydrogen/host.cs}
The Hamiltonian in~\autoref{eq:Hydrogen_Hamiltonian} is specified using the chemistry library in~\autoref{lst:specify-hamiltonian}.
The orbital integrals are represented by objects of the !OrbitalIntegral! class illustrated in~\autoref{line:orbital-integrals}.
A fermion Hamiltonian is represented by objects of the !FermionHamiltonian! class, and is constructed, as shown in~\autoref{line:construct-hamiltonian}, from the !OrbitalIntegral! array, together with the number of orbitals, and the number of electrons.
Note that the identity term $h_0$ is a constant energy offset, in this case representing Coulomb repulsion, and is also added to the Hamiltonian. 
The input state acted on by the Hamiltonian is represented by the objects of the !InputState! class.
When unspecified, $\ket{\psi}$ will default to a single Slater determinant by greedily minimizing the energy of diagonal one-electron terms.
For the case of Hydrogen, this state is a reasonable approximation of the ground state.
The effect of this optional step is reproduced in~\autoref{line:greedy-state-prep}.
If the hydrogen Hamiltonian is provided in the Broombridge schema, say the file !hydrogen.yaml!, the !FermionHamiltonian! instance may be more conveniently constructed as in~\autoref{line:create-from-broombridge}.
The schema is defined to contain at least equivalent information, so no other parameters have to be set.

As we target a qubit quantum computer, this fermion Hamiltonian must be converted into an equivalent Hamiltonian represented by qubit spin operators.
One possible representation of fermions is the Jordan--Wigner encoding~\autoref{sec:review_jw}.
This qubit Hamiltonian is represented by objects of the !JordanWignerEncoding! class, which can be easily obtained from !FermionHamiltonian! instances, as shown in~\autoref{line:convert-Jordan--Wigner}.
A final step in~\autoref{line:qsharp-format} is converting this \csharp~Hamiltonian object data structure into one that may be passed to \qsharp.
\lstinputlisting[
style=CSharp,
gobble=0,
linerange={77-110},
label={lst:specify-hamiltonian},
caption={Specifying a Hydrogen Hamiltonian in the chemistry library.\hfill\texttt{examples/hydrogen/host.cs}}
]{\sourcefolder/examples/hydrogen/host.cs}

\lstDeleteShortInline!

%% file: simulation_q_sharp.tex
The quantum simulation algorithms of Trotter--Suzuki and Qubitization are implemented in the \qsharp~canon library of the Microsoft Quantum Development Kit.
The chemistry library provides an interface to invoke these algorithms using the \csharp~Hamiltonian data structures of the previous section. 
In this section, we use the chemistry library to obtain operations representing preparation of the initial quantum state, and dynamical evolution by these Hamiltonians.
We start by importing the \qsharp~component of the chemistry library as follows.
\begin{lstlisting}[style=qsharp,label={lst:import-qsharp-chem-library},caption={Importing the Q\# component of the chemistry library.\hfill\texttt{examples/hydrogen/hydrogen.qs}}]
    // Imports Q# components of the chemistry library for the Jordan-Wigner encoding. 
    open Microsoft.Quantum.Chemistry.JordanWigner;  
    
    // Import the math constant //*{\color{comment-color}$\pi$}*// and the real modulus function.
    open Microsoft.Quantum.Extensions.Math
\end{lstlisting}

As the required number of qubits depends on the choice of quantum simulation algorithm, this information is also returned by the chemistry library. 
The results of these steps, illustrated in~\autoref{lst:obtain-operations}, are used to obtain eigenstate energy estimates in the next section. 
In~\autoref{line:unpack-qsharpdata}, we deconstruct the "qSharpData" data structure, representing hydrogen, from the previous section into components to be processed in \qsharp. 

We consider two simulation techniques.
First, in~\autoref{line:chemistry-trotterization}, we use the chemistry library to synthesize an operation that implements a single step of the Trotter--Suzuki integrator.
The integrator order is set in~\autoref{line:chemistry-trotterization-order}.
This circuit approximates real-time evolution $e^{-iHt}$, where $t$ is the integrator step size, set in~\autoref{line:chemistry-trotterization-stepsize}.
Second, in~\autoref{line:chemistry-qubitization}, we instead synthesize an operation that implements a quantum walk by $H$ using the Qubitization procedure.
This quantum walk implements the unitary with spectrum similar to $e^{i\Y\otimes sin^{-1}{(H/h)}}$, which may be understood as time-evolution by the Hamiltonian, but with a modified spectrum. 
Note that $h$ is the $L_1$ norm of the Hamiltonian term coefficients.
An operation that synthesizes the \lstinline[style=csharp]+``Greedy''+ input state, specified in the previous section is obtained in~\autoref{line:chemistry-state-preparation}.

\begin{lstlisting}[
style=QSharp,
label={lst:obtain-operations},
caption={Quantum simulation circuits by the Hamiltonian in~ \autoref{lst:specify-hamiltonian}.\hfill\texttt{examples/hydrogen/hydrogen.qs}}
]
    //The  Q# data structure `qSharpData` from //*\autoref{line:qsharp-format} of~\autoref{lst:specify-hamiltonian}*// represents hydrogen.
    // This deconstructs into variables that will be passed to later Q# algorithms.
    // `nSpinOrbitals` is the number of spin-orbitals.
    // `data` describes Hamiltonian terms.
    // `statePrepData` describes the input state the simulation algorithm acts on.
    // `energyOffset` is the identity coefficient of the Hamiltonian.
    let (nSpinOrbitals, data, statePrepData, energyOffset) = qSharpData!;//*\label{line:unpack-qsharpdata}*//

    // Using Trotterization ////////////////////////////////////////////////////*\label{line:chemistry-trotterization}*//
    // Set the Trotter--Suzuki integration order.
    let order = 1;//*\label{line:chemistry-trotterization-order}*//
    // Set the Trotter--Suzuki integration step size.
    let stepSize = 0.4;//*\label{line:chemistry-trotterization-stepsize}*//
	
    // Apply the chemistry library function `TrotterStepOracle` to the Q# 
    // data structure representating hydrogen.
    // `nQubits` is the number of qubits allocated to the quantum circuit.
    // `rescale` is `1.0/trotterStepSize`, and is the number of Trotter steps
    // needed to simulation evolution for unit time.
    // `trotterStep` is an operation implenting one Trotter--Suzuki step.
    let (nQubits, (rescale, trotterStep)) = 
        TrotterStepOracle(qSharpData, stepSize, order);
	
    // Using Qubitization //////////////////////////////////////////////////////*\label{line:chemistry-qubitization}*//
    // Apply the chemistry library function `QubitizationOracle` to the Q# 
    // data structure representating hydrogen.
    // `nQubits` is the number of qubits allocated to the quantum circuit.
    // `rescale` is the //*{\color{comment-color}$L_1$}*// norm of Hamiltonian coefficients.
    // `quantumWalk` is an operation implenting one quantum walk step.
    let (nQubits, (l1Norm, quantumWalk)) = QubitizationOracle(qSharpData);
    
    // State Preparation ///////////////////////////////////////////////////////*\label{line:chemistry-state-preparation}*//
    // Apply the chemistry library operation `PrepareTrialState` to 
    // synthesize a quantum circuit that prepares the initial state acted on by
    // the Hamiltonian. Note that the partial application of the last paramter, 
    // which has the `Qubit[]` type.
    let statePrep =  PrepareTrialState(statePrepData, _);
\end{lstlisting}

%% file: simulation_estimate_energy.tex
The operations we assembled in the previous section simulate quantum dynamics.
These operations may be combined with phase estimation---described in~\autoref{sec:phase_estimation}---to estimate energy levels $E_n$ of the Hamiltonian.
If the code from~\autoref{lst:obtain-operations} is used to specify the approximation to $e^{-iHt}$ and the initial state, then the application of phase estimation here will provide an estimate of the ground-state energy of hydrogen with high probability because of the large overlap between the ground-state of hydrogen and the Hartree--Fock trial state used therein.

Below in~\autoref{lst:obtain-energy}, we illustrate this procedure by combining the chemistry library with phase estimation in the canon library.
Specifically, we apply robust phase estimation below to estimate an eigenvalue of the Hamiltonian.
\begin{lstlisting}[
style=QSharp,
label={lst:obtain-energy},
caption={Obtaining an estimate of an eigenvalue of the Hamiltonian.\hfill\texttt{examples/hydrogen/hydrogen.qs}}
]
    // Choose robust phase estimation amongst various possible phase 
    // estimation algorithms in the Q# canon. This is parameterized by bits
    // of precision `b` which controls the standard deviation $\approx 2^{b-1}$ of
    // the returned phase. 
    let nBits = 8;
    let phaseEstAlg = RobustPhaseEstimation(nBits, _, _);	
	
    // Use the Q# canon function `EstimateEnergy` that allocates qubits
    // to the quantum simulation, creates the input states, performs 
    // phase estimation, and then returns the estimated phase. 
    let phaseEst = EstimateEnergy(nQubits, statePrep, trotterStep, phaseEstAlg);
	
    // Rescale the estimated phase to obtain the energy, and also add the 
    // identity coefficient of the Hamiltonian.
    let energyEst = phaseEst * rescale + energyOffset;
	
    // Suppose a guess of the true energy is available to identify the principle value
    // range.
    let energyGuess = -1.0;
	
    // Add the multiples of $2\pi$ to obtain the principle value using
    // the canon `RealMod` modulus function.
    let principleEnergyEst = 
       RealMod(energyEst, 2 * PI() * rescale, energyGuess - PI() * rescale);
\end{lstlisting}

This code is quite modular.
It is easy to replace the Trotter--Suzuki simulation algorithm used here with another unitary operation, such as those yielded by qubitization or linear combinations of unitaries.
It is also worth noting that, while the initial state preparation provided previously will tend to have a high-overlap with the ground state for this problem, such an elementary ansatz may not be appropriate in some cases.
In examples of the next section, we will generalize this state to probe the excited states of molecules, which is often a more challenging problem than finding ground-state energies for classical computers.

%% file: applications.tex

In this section, we illustrate our quantum chemistry library and interoperation with  NWChem by studying several model systems that epitomize typical challenges encountered in realistic molecular simulations.
Evaluating the behavior of Trotter--Suzuki simulation algorithms integrators from~\autoref{sec:review_ts} in conjunction with phase estimation algorithms from~\autoref{sec:phase_estimation} for situations characterized by high levels of electronic wavefunction quasi-degeneracy is of special importance in these investigations.
Such situations naturally occur in studies of ground-state potential energy surfaces, bond-forming and bond-breaking processes, as well as condensed matter systems approaching metallic regimes. 
These problems usually elude standard formulations, especially in cases when one cannot define proper reference function or the use of multi-reference concepts and model/active spaces leads to the emergence of intruder state problems \cite{schucan1972effective,hose1979diagrammatic,zarrabian1990applicability,finley1995applications}.

We model systems broadly used in testing and verification of high-accuracy methods that exhibit such quasi-degeneracies.
First, we study potential energy surfaces of LiH for the ground state and excited states, which are typically strongly varying with a large doubly excited component.
We show how to build these models for the Hamiltonian and states of LiH and simulate them through our reproducible workflow.
By using elementary coupled-cluster ansatzes, which are conveniently represented using Broombridge, the excited states of LiH are probed for a range of different internuclear separations and the success probability for this is found to typically quite large.  
These results not only illustrate the utility of quantum chemistry simulation library, but suggests that understanding the excited states of molecules will be an important application for quantum computers.

Second, we rigorously evaluate the ground-state energy of stretched H$_{10}$. This system is typically used to model situations where almost all orbitals need to be considered as active, especially for the large H--H separations. 
An extensive discussion of the H$_{10}$ system properties can be found in Ref.~\cite{motta2017towards}.
Whereas the ground-state energy of H$_{10}$ is challenging to describe even using high-order coupled-cluster methods, we show that quantum computing is capable of providing accurate ground-state energy estimates.
This example also performs sweeps, across different Trotter--Suzuki step-sizes and the precision of phase estimation to mimic how one might, on quantum hardware, empirically verify that chemical accuracy has been reached.
Such sweeps would be difficult to do by hand as dozens of configurations need to be probed in these studies, which shows that such software allows for qualitatively different types of research that would be inconvenient without it.

Third, we compare the ground-state energy of different C$_{20}$ isomerizations by employing an active space.
Even more accurate results can be obtained by a larger $50$-orbital active space. 
While these large examples are too difficult to simulate on classical computers, we are capable of using our software to automatically estimate the number of gates required for a simulation.
We further find that the cost of such simulations is comparable to previous estimates generated for FeMoco~\cite{Reiher2016Reaction}, but without requiring any form of extrapolation from empirical results.
This provides evidence suggesting that the applied algorithm, an optimized implementation of qubitization, may be a favored method for simulating challenging problems on fault-tolerant quantum computers.

\subsection{Lithium hydride ground- and excited-state energies}
\label{sec:example_LiH}
Studies of excited-state potential energy surfaces rely on good-quality initial choices for the approximate excited states that provide a reasonable overlap with the exact excited states. 
An illustration is provided by identifying several lowest-lying excited states of the LiH molecule with $12$ spin orbitals in the STO-3G basis set as a function of the internuclear Li-H distance $R_{\rm Li\text{-}H}$.
In \autoref{tab:table_r}, we collate the leading coefficients of the excited-state wavefunction expansion corresponding  to the largest  EOMCCSD amplitudes defining   $R_{K,1}$ and $R_{K,2}$ operators defined in~\autoref{eq:excc}.
As one can see from the table, even for close-to-equilibrium geometry $R_{\rm Li\text{-}H}=1.6$\AA, all five lowest-lying singlet excited states reveal multi-configurational character where several excited Slater determinants play key roles. 

These excited states provide a good illustration of the importance of double excitations.
As can be seen from \autoref{tab:table_r}, even the first excited state, dominated by a single excitation, has non-negligible contributions from doubly excited Slater determinants.
More precisely, let $\ket{\Psi_\mathrm{HF}}$ be the single-configuration Hartree--Fock state.
Then the first excited state $\ket{\Psi_{\mathrm{E1}}}$ ansatz close to the equilibrium geometry is
\begin{align}
	\ket{\Psi_{\mathrm{E1}}}\propto\left(
	0.889 (a^\dagger_{3,\uparrow}a_{2,\uparrow}+a^\dagger_{3,\downarrow}a_{2,\downarrow})
	+0.221(a^\dagger_{3,\uparrow}a^\dagger_{6,\downarrow}+a^\dagger_{6,\uparrow}a^\dagger_{3,\downarrow}) a_{2,\uparrow}a_{2,\downarrow}
	-0.324a^\dagger_{3,\uparrow}a^\dagger_{3,\downarrow} a_{2,\uparrow}a_{2,\downarrow}
	\right)\ket{\Psi_\mathrm{HF}}.
\end{align}
Note that the produced state is always correctly normalized, meaning  $\braket{\Psi_{\mathrm{E1}}|\Psi_{\mathrm{E1}}}=1$, by rescaling the input coefficients if necessary.

The situation becomes more complicated when the Li-H distances are stretched.
For example, at $R_{\rm LiH}=1.6$\AA, all states acquire mixed single and doubly excited character.
If an elementary single excitations Hartree--Fock state ansatz is used, then the success probability for phase estimation is unlikely to be high for excited states (especially as the inter-nuclear distance grows). 
This illustrates the importance of having a reliable many-body framework to provide a plausible initial guess for excited-state simulations using phase estimation algorithms.
\setlength{\tabcolsep}{6pt}
\begin{center}
	\begin{table*}
		\centering
		\begin{tabularx}{\textwidth}{clll} \hline \hline
		$R_{\rm Li-H}$	& 
		Excitation energy (eV)	& $R_{1}$ & 
		$R_{2}$    \\
		\hline 
		1.600 &     3.618    &   $r^{2}_{3}=r^{\bar{2}}_{\bar{3}}=0.889$    &  $r^{2\bar{2}}_{3\bar{6}}= r^{2\bar{2}}_{6\bar{3}}= 0.221,\; r^{2\bar{2}}_{3\bar{3}}=-0.324$    \\[0.02cm]
				&     5.039    &   $r^{2}_{4}=r^{\bar{2}}_{\bar{4}}=0.707 ,\; r^{2}_{5}=r^{\bar{2}}_{\bar{5}}=0.542$  &  $r^{2\bar{2}}_{4\bar{6}}=r^{2\bar{2}}_{6\bar{4}}=0.204$  \\[0.02cm]
				&     5.039    &   $r^{2}_{5}=r^{\bar{2}}_{\bar{5}}=0.875$   &   $r^{2\bar{2}}_{5\bar{6}}=r^{2\bar{2}}_{6\bar{5}}=0.252$    \\[0.02cm]
				&    15.342   &   $r^{2}_{6}=r^{\bar{2}}_{\bar{6}}=0.325$    & $r^{2\bar{2}}_{3\bar{3}}=-0.595,\;  r^{2\bar{2}}_{3\bar{6}}=r^{2\bar{2}}_{6\bar{3}}=0.310,\;r^{2\bar{2}}_{4\bar{4}}=r^{2\bar{2}}_{5\bar{5}}=0.397 $    \\[0.02cm]
				&    17.947   &    $r^{2}_{3}=r^{\bar{2}}_{\bar{3}}=-0.775$   &    $r^{2\bar{2}}_{3\bar{3}}=-0.514,\; r^{2\bar{2}}_{3\bar{6}}=r^{2\bar{2}}_{6\bar{3}}=0.249$   \\[0.02cm]
		\hline
		4.000 &    2.239    &   $r^2_3=r^{\bar{2}}_{\bar{3}}=0.336,\;r^2_6=r^{\bar{2}}_{\bar{6}}=-0.477$   &   $r^{2\bar{2}}_{6\bar{6}}=0.236 ,\;r^{2\bar{2}}_{3\bar{6}}=r^{2\bar{2}}_{6\bar{3}}=0.353\;,r^{2\bar{2}}_{3\bar{3}}=-0.595$   \\[0.02cm]
				&    2.360    &    $r^2_5=r^{\bar{2}}_{\bar{5}}=0.602$  &   $r^{2\bar{2}}_{ 5\bar{6}}= r^{2\bar{2}}_{ 6\bar{5}}= -0.148,\;r^{2\bar{2}}_{ 3\bar{5}}= r^{2\bar{2}}_{ 5\bar{3}}=-0.545$  \\[0.02cm]
				&    2.360    &    $r^2_4=r^{\bar{2}}_{\bar{4}}=-0.602$   &  $r^{2\bar{2}}_{4\bar{6}}= r^{2\bar{2}}_{6\bar{4}}=0.148,\; r^{2\bar{2}}_{3\bar{4}}= r^{2\bar{2}}_{4\bar{3}}=0.545$   \\[0.02cm]
				&    8.023    &   $r^2_3=r^{\bar{2}}_{\bar{3}}=-0.102$    & $r^{2\bar{2}}_{6\bar{6}}=-0.219,\; r^{2\bar{2}}_{6\bar{3}}=0.341,\;r^{2\bar{2}}_{3\bar{3}}=0.605,\;r^{2\bar{2}}_{3\bar{6}}=-0.675$      \\[0.02cm]
				&    8.070    &  $r^2_4=r^{\bar{2}}_{\bar{4}}=-0.106$     & $r^{2\bar{2}}_{4\bar{3}}=-0.116,\; r^{2\bar{2}}_{6\bar{4}}=0.317,\;  r^{2\bar{2}}_{3\bar{4}}=0.934 $      \\[0.02cm]
		\hline 
		\hline\end{tabularx}
\caption{
	Leading EOMCCSD amplitudes of excited state, as described in~\autoref{eq:excc}, with absolute values greater than $0.2$ for two geometries of the LiH system in the STO-3G basis set.
	We use the orbital convention for denoting excitation amplitudes. 
	For example $r^{2}_{3}$ designates an excitation process of an $\alpha$ electron from orbital $2$ to $\alpha$ electron in orbital $3$, 
	$r^{\bar{2}}_{\bar{3}}$ designates an excitation process of a $\beta$ electron from orbital 2 to $\beta$ electron in orbital $3$
	TCE normalization of $R_K$ vectors are used.
	All distances are reported in Angstroms.
}
		\label{tab:table_r}
	\end{table*}
\end{center}

This example highlights the ability to represent multi-configurational wavefunctions in Broombridge, and to prepare these quantum states on a quantum computer.
By using trial wavefunctions with sufficient overlap with the desired excited state, we may then obtain a targeted sweep over excited-state energies.
Higher excitations, such as triple and quadruple excitations, are needed in many situations to accurately describe excited-state potential surfaces and corresponding topological events including minima, avoided crossings, and conical intersections.
These excitations may be similarly represented in Broombridge, though note that the cost of state preparation scales polynomially with the number of configurations.

\begin{figure}
	\centering
	\includegraphics[width=1.0\linewidth]{\figurefolder/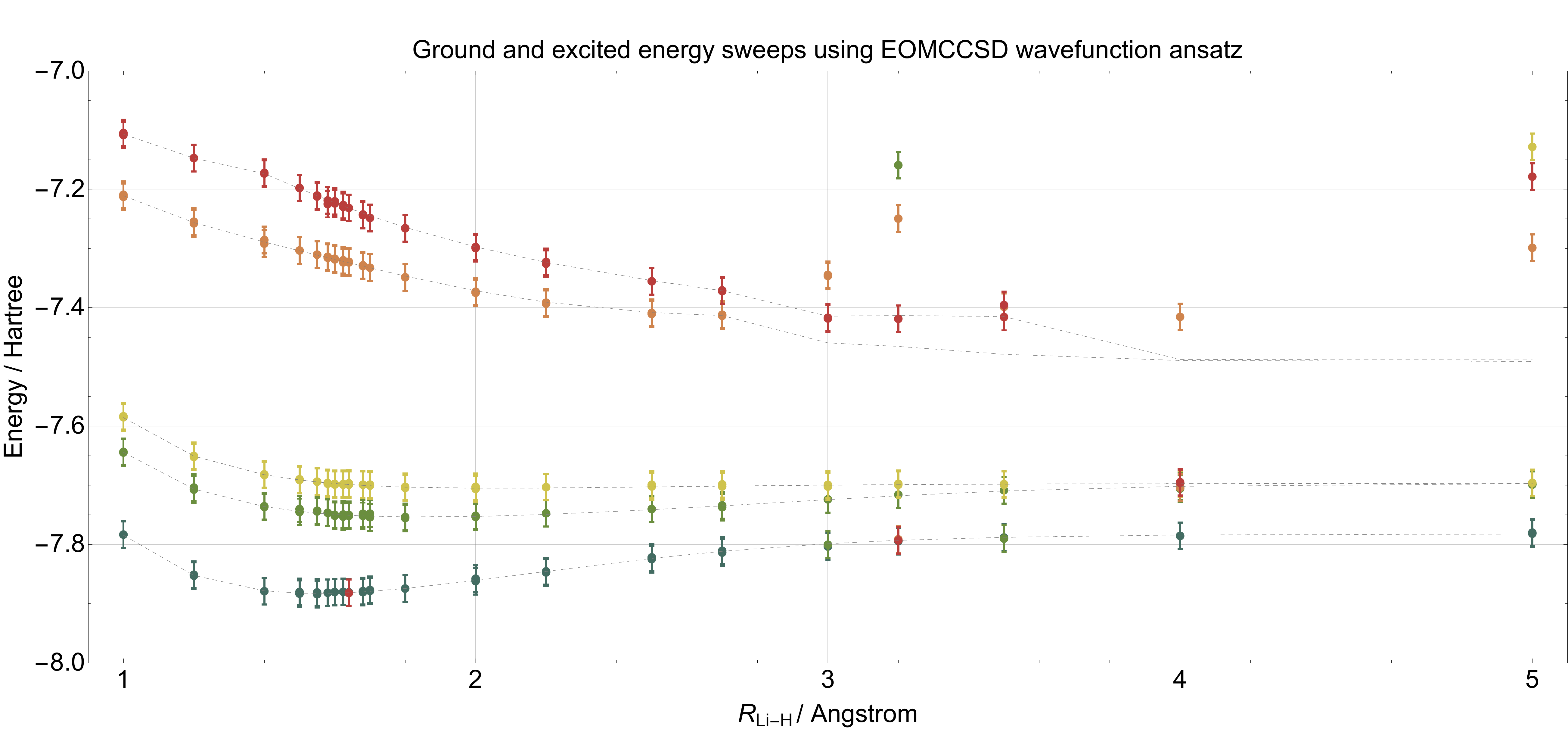}	
	\caption{Estimated eigenstate energies as a function of $R_{\rm{Li\text{-}H}}$ bonding distance obtained using robust phase estimation with a Trotter step size of $t=0.5$, $b=10$ bits of precision, and two repetitions at each bonding distance per energy level.
	Error bars correspond to contribution of the phase estimation procedure alone, and are exaggerated by a factor of $5$ for clarity.
	Exact FCI energies are plotted (gray) for comparison.
	Different colored points correspond to initial state ansatzes targeted at different energy levels. Each ansatz is multi-configurational with up to the first $12$ dominant coefficients included.
	This enables a selective sweep of the desired excited-state energy, though observe that at large bond distances, particularly the fourth excited state, the ansatz may still have a low probability of projecting onto the exact eigenstate. }
	\label{fig:LiH}
\end{figure}

In~\autoref{fig:LiH}, we plot the results of LiH eigenstate energy estimation.
Each point-energy estimate is obtained through the following steps.
\begin{enumerate}[noitemsep,nolistsep]
	\item NWChem is used to generate a description of the electronic structure problem at the desired bond distance, following the example of~\autoref{sec:review_nwchem}.
	\item We generate a Broombridge representation of LiH from the NWChem output following~\autoref{sec:broombridge_generation}.
	\item Each LiH instance represented in Broombridge is imported, following~\autoref{line:create-from-broombridge} of ~\autoref{lst:specify-hamiltonian}.
	\item We select the initial state ansatz to be used in each LiH simulation, following~\autoref{line:qsharp-format} of ~\autoref{lst:specify-hamiltonian}.
	\item Finally, we execute robust phase estimation, following~\autoref{lst:obtain-energy}, using a Trotter step size of $t=0.5$, and $b=10$ bits of precision. 
\end{enumerate}
Note that with these choices of parameters, where the Trotter step-size is chosen a posteriori, the empirical error of robust phase estimation is $t/2^{b-1}\approx0.00098$, which is sufficient for chemical accuracy of $10^{-3}$, ignoring simulation errors from the finite Trotter step size.
Empirically evaluating the Trotter step size will be the subject of the next example.

\subsection{Empirical Trotter error estimation of stretched \texorpdfstring{H$_{10}$}{H10} chains}

The H$_{10}$ system epitomizes many of the correlation effects encountered in realistic strongly correlated molecular systems.  
This system has been recently used in studies of various methods designed to deal with strong quasi-degeneracy effects~\cite{hachmann2006multireference} as the degree of multi-configurational character of the ground-state wavefunction can be varied by changing the hydrogen-hydrogen distance $R$ in a linear chain of the hydrogen atoms.
For example, at $R=1.0$~a.u., the corresponding wavefunction is dominated by a single restricted Hartree--Fock  determinant.
However stretching the $R$ distance to $3.6$~a.u.~results in a multi-reference character of the wavefunction, which poses a significant challenge for single reference CC methods using RHF reference. Among several methods tested in Ref.~\cite{motta2017towards}, the auxiliary-field quantum Monte Carlo (AFQMC), density matrix embedding theory (DMET), UHF-CC, and self-energy embedding theory (SEET) formalisms lent themselves to coping with strong correlation effects.
Other formalisms such as self-consistent second-order Green's function (GF2) and RHF-CC approximations suffer significant deterioration in energy accuracies especially for stretched geometries.
For this reason, the H$_{10}$ system is an ideal target for testing various elements of quantum algorithms for electronic structure problems.
We also compare and discuss the accuracies of state-selective multi-reference coupled-cluster (MRCC) methods, represented here by the Mukherjee's MRCC approach (MkMRCC)~\cite{mahapatra1999size,mahapatra1999development}, which was not studied in Ref.~\cite{motta2017towards}.

We perform quantum simulations within an STO-6G minimal basis consisting of $20$ spin orbitals. 
The results of robust phase estimation on a first-order Trotter-Suzuki formula, similar to LiH in the previous section, are compared to various classical approaches.
The MkMRCC formalism has been tested using its two variants: MkMRCC model with singles and doubles (MkMRCCSD) and MkMRCCSD with perturbative triples corrections (MkMRCCSD(T)); see~\cite{bhaskaran2008multireference,bhaskaran2011multireference,bhaskaran2012implementation}.
In~\autoref{tab:h10}, we have also collated CCSDT~\cite{ccsdt_noga,ccsdt_noga_err,scuseria_ccsdt} and CCSDTQ~\cite{Kucharski1991,ccsdtq_nevin} ground-state energies, which also have not been studied in Ref.~\cite{motta2017towards}.

As a result of strong quasi-degeneracy effects, both CCSD and CCSDT energies start to significantly deviate from their FCI counterparts for stretched geometries. 
Although the non-variational collapse of the CCSDT energies is not as profound as in the CCSD case, we notice a sizable CCSDT energy error of $47$ milli\hartree~at $R = 3.6$ a.u.
The addition of  quadruple excitations in the CCSDTQ method offsets the variational collapse of the CCSD method. However, a $4$ milli\hartree~still persists for larger distances.

Although resorting to the MkMRCCSD formalism significantly improves the quality of energies in the $R=1.0$~a.u. to $R=3.2$~a.u. region, the MkMRCCSD approach still yields large negative errors at $R = 3.6$ a.u..
This situation is mostly a consequence of an inadequate choice of model space and the resulting intruder state problem~\cite{evangelisti1987qualitative}.
These are manifestations of divergent perturbation theory expansions that occur in near-degenerate systems and are the culprits behind the divergent character of MkMRCCSD(T) correction.

Phase estimation approaches on quantum computers remove the biases introduced by the use of reference function, model space, and level of excitation.
In contrast to other results collated in~\autoref{tab:h10}, phase estimation produces errors of small uniform size for all geometries.
This is especially important for stretched geometries, particularly $R=3.6$~a.u., where the CCSD, CCSDT, MkMRCCSD, and MkMRCCSD(T) approaches reveal singular behavior.
Even though the failure of the CCSDTQ approach is not as pronounced as it is in the other cases, the CCSDTQ energy error of $4.5$~milli\hartree~for $R = 3.6$~a.u. is significantly larger than the $0.7$~milli\hartree~error obtained with phase estimation.
In principle, the error of phase estimation can be made arbitrarily small, at a modest proportionately higher cost.
This presents a strong argument in favor of using quantum simulation algorithms within quasi-degenerate or metallic regimes. 

\begin{figure}
	
	\includegraphics[width=1.0\linewidth]{\figurefolder/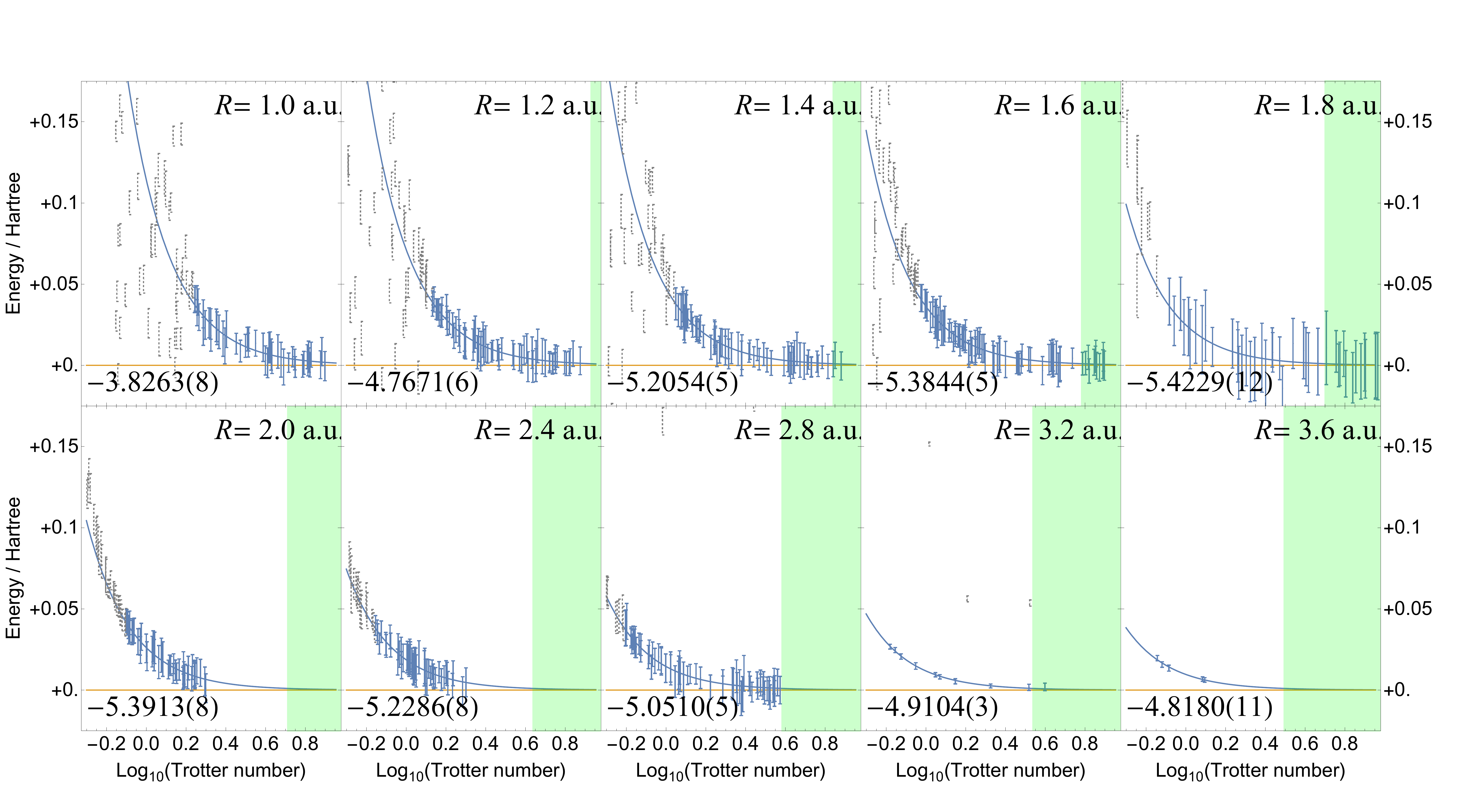}
	\caption{Estimated ground-state energy $E$ (orange) as a function of Trotter number $r$ of the first-order integrator $\prod_{j} e^{-iH_j/r}$ for H$_{10}$ at various bond distances obtained by a least-squares fit to $E=\frac{m}{r^2}+E_0$ (blue), where $m,E_0$ are fit parameters.
	The ground-state energy estimate $E_0$, which is assumed to be normal distributed, is also plotted (yellow) with its standard deviation provided in round brackets.
	Error bars represent the standard deviation of $E_0$ by robust phase estimation, given the prior that $E_0$ lies in the plotted range.
	Dotted error bars (gray) are excluded from the fit as they either have a large Trotter--Suzuki error, or correspond to excited states.
	The shaded region (green) corresponds to a Trotter number that achieves at least chemical accuracy based on the fit $\frac{m}{r^2}\le 0.001$~\hartree.}
	\label{fig:h101}
\end{figure}

In these examples, the Trotter number chosen for the Trotter--Suzuki formula, which is the inverse of the Trotter step-size as described after~\autoref{eq:trotter_first_order}, in the phase estimation algorithm can be found in~\autoref{fig:h101}.
We see that, despite the errors in all the coupled-cluster methods considered in~\autoref{tab:h10} that increasing with $R$, the Trotter number needed to achieve chemical accuracy actually decreases monotonically with $R$, which corresponds to a more efficient quantum simulation.
This increase is unsurprising as the Coulomb interaction between electrons, which typically dominates errors in the quantum simulation algorithms, weaken when the molecule stretches.

A major strength arising from the integration of \qsharp~and NWChem is the ability to automate these sweeps of the FCI energy as a function of $R$.
This automation means that studies of phase estimation over different spacings in H$_{10}$, or surveys of hydrogen chains of different lengths, can be easily generated inside a single framework.
Furthermore, all code written for these simulations will still be useful once quantum computers are built that can reliably perform phase estimation on H$_{10}$ because of the simulator model used within \qsharp~forbids the language from distinguishing between the simulator and a quantum computer.  
This work provides a preview of the types of studies that automated quantum computer simulations of chemistry will one day enable while illustrating the challenges that can be faced by classical methods for even simple molecules.

\begin{center}
    \begin{table*}
        \centering
        \begin{tabularx}{\textwidth}{ XX|cccccc}
            \hline\hline
            \multirow{2}{*}{R (a.u.)}&      \multirow{2}{*}{FCI (\hartree)}    &   \multicolumn{6}{c}{Difference from FCI ($10^{-3}$ Hartree)}   \\
            && CCSD  &  CCSDT  &  CCSDTQ  &  MkMRCCSD  &  MkMRCCSD(T) & RPE   \\
            \hline
			1.0& -3.82439&   0.5&   0.0&   0.0&   0.1&  -0.1&   -1.9 $\pm$ 0.8\\ 
			1.2& -4.76638&   0.8&   0.0&   0.0&   0.0&  -0.2&   -0.7 $\pm$ 0.6\\ 
			1.4& -5.20509&   1.1&  -0.1&   0.0&   0.0&  -0.5&   -0.3 $\pm$ 0.5\\ 
			1.6& -5.38436&   1.4&  -0.2&   0.0&  -0.1&  -0.8&   0.0 $\pm$ 0.5\\ 
			1.8& -5.42439&   1.8&  -0.5&   0.0&  -0.3&  -1.4&   1.5 $\pm$ 1.2\\ 
			2.0& -5.38963&   2.2&  -1.1&   0.0&  -0.4&  -2.2&   -1.7 $\pm$ 0.8\\ 
			2.4& -5.22794&   1.1&  -5.6&   0.0&   0.3&  -4.0&   -0.7 $\pm$ 0.8\\ 
			2.8& -5.05024& -17.1& -32.7&   0.6&   0.6&  -8.9&  -0.8 $\pm$ 0.5\\ 
			3.2& -4.91038& -990.0& -888.0&   3.5&  -1.3& -19.4&  -0.4 $\pm$ 0.4\\ 
			3.6& -4.81870& -111.0& -47.3&   4.5& -43.3 $\pm$ 1.0& -74.3 $\pm$ 1.0& 0.7 $\pm$ 1.1 \\
            \hline \hline
        \end{tabularx}
		\caption{
            Comparison of energies with respect to bond distance $R$ obtained by phase estimation through the Microsoft Quantum Development Kit Robust Phase Estimation (RPE) algorithm with energies obtained by various CC methodologies for the H$_{10}$ system in the STO-6G basis set. In all CC calculations, restricted Hartree--Fock molecular orbitals are used.
        }
        \label{tab:h10}

    \end{table*}        
\end{center}

In~\autoref{fig:h101}, we plot the results of H$_{10}$ eigenstate energy estimation.
Each point-energy estimate is obtained as follows.
\begin{enumerate}[noitemsep,nolistsep]
	\item NWChem is used to generate a description of the electronic structure problem at the desired bond distance, following~\autoref{sec:review_nwchem}.
	\item We generate a Broombridge representation of H$_{10}$ from the NWChem output following~\autoref{sec:broombridge_generation}.
	\item Each H$_{10}$ instance represented in Broombridge is imported, following~\autoref{line:create-from-broombridge} of ~\autoref{lst:specify-hamiltonian}.
	\item Select the ground-state ansatz in Broombridge, following~\autoref{line:qsharp-format} of ~\autoref{lst:specify-hamiltonian}.
	\item Execute robust phase estimation, following~\autoref{lst:obtain-energy}, using a Trotter step size $t\in[t_{\rm{min}},t_{\rm{max}}]$ uniformly sampled on a log scale, and some $b$ bits of precision.
\end{enumerate}
In contrast to~\autoref{sec:example_LiH}, where the Trotter number $r=1/t$ and bits of precision $b$ were chosen a posteriori, we empirically determine the Trotter step-size $t$ required to control the error contribution  $\Delta$ from the Trotter--Suzuki integrator.
We do so by applying the known error scaling  from~\autoref{eq:trotter_first_order}
\begin{align}
	\left\|e^{-iHt} - \prod_{j=1}^M e^{-ih_j P_j t}\right\| = \mathcal{O}(t^2)
\end{align}
for the first-order integrator that is used here.
Thus, in the limit of sufficient small $t$, the estimated ground state energy $E$ ought to scale as 
\begin{align}
	E= E_0 + \Delta = E_0 + m t^2 +\mathcal{O}(t^3),
\end{align}
where $E_0$ is the exact ground-state energy and $m$ is a constant.
Thus, performing phase estimation on $E$ for various step sizes $t$ allows us to obtain data that we fit to obtain $E_0$ and $m$, in the regime where the quadratic $t^2$ scaling is observed to be dominant.
Note that by repeating $\mathcal{O}(1/\Delta^2)$ times, achieving chemical accuracy is possible through the fit, even if the sampling error from a single run of phase estimation is large.
However, overall quantum gate complexity is always minimized by performing a logarithmic number of high-precision estimates which takes $\mathcal{O}(1/\Delta)$ time, rather than the $\mathcal{O}(1/\Delta^2)$ time required by na\"{\i}ve sampling.
In particular, the fitted energy estimates can be highly accurate even if all individual data points have both a Trotter--Suzuki approximation error and a phase estimation error larger than chemical accuracy.

\subsection{Quantum computing predictions and resource estimations for \texorpdfstring{C$_{20}$}{C20} isomerizations}

The relative energies of the three C$_{20}$ fullerene, bowl, and ring isomers have been a subject of intensive theoretical studies spanning nearly three decades~\cite{martin1996structure,brabec1992precursors,zhang1992geometry,raghavachari1993isomers,von1993carbon,
grossman1995structure,taylor1995c20,bylaska1996lda,galli1998tight,sokolova2000energetics,grimme2002structural,an2005ab,jin2015coupled,
apra2016implementation}.
Although these three isomers have been widely examined both experimentally and theoretically, there is still controversy regarding their relative stability.
The magnitude of disagreement between different electronic structure theories is particularly surprising.
Even though these differences were mapped out in the early 1990s---and despite many subsequent calculations sometimes in dispraise of the original calculations---the energetic orderings and nature of the differences between the different theories are still unknown.

Several HF DFT-generalized gradient approximations (GGA) electronic structure methods using various basis sets predict the ring isomer energy to be more than $2$ eV lower than the fullerene isomer, with the bowl isomer somewhere in between.
However, DFT-local density approximation (LDA), MP2, and CCSD(T) methods predict the opposite trend: the fullerene and ring are now the lowest and highest energy isomers; the fullerene isomer is at least $1.7$ eV lower in energy than the ring.
Adding even more uncertainty to the reliability of electronic structure methods for this system is the fact that calculations carried out using the diffusion Monte-Carlo (DMC) method predict a completely different ordering:  the bowl is lower in energy from the ring and fullerene by $1$ eV and $2$ eV respectively.
These dramatic swings in relative energies of up to $4$ eV between the two best methods---CCSD(T) and DMC---for treating electron correlation
are quite shocking, but not completely surprising since attaining the correct energetics for molecules containing both delocalized $\pi$-bonding and carbon-carbon triple bonds is still a challenge for electronic structure methods.
 
The widely varying results of earlier simulations clearly indicate that proper inclusion of electron correlation effects plays a key role in establishing the proper energetic ordering of these systems.
The relative stability of the isomers is quite sensitive to the method and basis set choice, and quantum computing may play an important role in describing these systems in the future.
To demonstrate this feasibility, we consider phase estimation calculations using active spaces that begin to capture both the extended bonding and triple bonds in the molecules.
Due to the limitations of current quantum computing platforms, we can only perform phase estimation simulations in relatively small active spaces.
For this reason, we focus our attention on fullerene and ring C$_{20}$ isomers, where---in contrast to the bowl isomer---the small active spaces composed of eight electrons distributed among eight orbitals have the potential to  capture essential correlations effects.

We performed simulations for C$_{20}$ using phase estimation on a Trotter--Suzuki formula in the cc-pVDZ basis set and employing fullerene and ring C$_{20}$ geometries utilized by~\cite{apra2016implementation}.
For evaluating fullerene and ring C$_{20}$ energies, we used the first $150$ energy estimates that were not discarded for either corresponding to an excited-state energy, or begin an outlier based on a two-sided Grub's test with a confidence level of 0.05.
The fullerene-ring energy differences are shown in~\autoref{table_c20} in units of eV and are in qualitative agreement with earlier simulations.
For example, the separation of $1.77$ eV as obtained with the Microsoft Quantum Development Kit is very close to the $1.7$ eV obtained with the CCSD(T) formalism in \citet{bylaska1996lda}.
Note that the workflow per obtaining each energy estimate closely mirrors the LiH and H$_{10}$ examples previously discussed.
\begin{center}
	\begin{table*}
		\centering
		\begin{tabularx}\textwidth{XXXXXXXX} \hline \hline
			Algorithm &
			MkCCSD & 
			MkCCSD(T)  &
			CCSD(T) &
			CCSD(T) &
			CCSD(T) &
			CCSD(T) &
			RPE \\[0.02cm]
			\hline
			Energy (eV) & 1.15 & 2.03 & 2.2 & 1.7 & 2.06 & 1.04 & 1.77   \\[0.02cm]
			\hline\hline
		\end{tabularx}
		\label{table_c20}
		\caption{The energetic separation between fullerene and ring  C$_{20}$ configurations.  
			Microsoft Quantum Development Kit Robust Phase Estimation (RPE) algorithm results were obtained with an active space composed of eight electrons distributed among eight molecular orbitals using cc-pVDZ basis set.
			All separations are reported in electron volts. The Mk-MRCCSD and Mk-MRCCSD(T) results were taken from Ref.~\cite{apra2016implementation}.
			The cc-pVDZ MP2/CCSD(T) results for SCF geometries were taken from Ref.~\cite{bylaska1996lda}.
			 The cc-pVDZ CCSD(T) results for the LDA geometries were taken from Ref.~\cite{bylaska1996lda}.
			 The cc-pVDZ CCSD(T) result of Ref.~\cite{an2005ab}. 
			 The cc-pVTZ CCSD(T) (drop core) results of Ref.~\cite{jin2015coupled}.\\}
	\end{table*}
\end{center}

Although, small-size active-space type simulations for C$_{20}$ fullerene and ring configurations using the Microsoft Quantum Development Kit corroborate previous predictions of large CC calculations, one should take these results with caution.
In the active-space Microsoft Quantum Development Kit simulations, we eliminate a large number of occupied orbitals.
Therefore, the one- and two-electron integrals for the core Hamiltonian and Coulomb interactions limited to active-space molecular indices, leading to a form  of the Fock matrix that is different from the active-active block of the full Fock matrix that includes summation over two-electron integrals involving  non-active occupied molecular indices.
This is the major reason why one obtains the opposite energy ordering in correlated CC calculations with all but active orbital frozen, yet employing active-active block of the full Fock matrix.
For example, our NWChem CCSDTQ calculations correlating $8$ electrons in $8$ orbitals place the ring isomer energy $1.23$ eV below the fullerene isomer energy.
For consistency, we performed CCSDTQ calculations using integrals convention employed in Microsoft Quantum Development Kit active-space calculation which resulted in CCSDTQ energies very close to ones obtained in Microsoft Quantum Development Kit simulations.
This disagreement may be resolved either through Microsoft Quantum Development Kit simulations employing particle-hole representation of creation/annihilation operators~\cite{Barkoutsos2018particlehole} or by employing larger active spaces, as discussed next.

\subsubsection{Resource estimation for simulating \texorpdfstring{C$_{20}$}{C20} isomers}


Until fault-tolerant quantum computers are available, classical simulations of quantum chemistry at chemical accuracy are limited to $40-50$ spin orbitals.
Nevertheless, we may still obtain estimates of the quantum resources required to execute these classically intractable examples.
Resource estimates are tabulated in~\autoref{tab:c20} for two implementations of qubitization: one unoptimized and the other optimized for chemistry as described in~\autoref{sec:optimized_qubitization}.
Data in each row are obtained by the following procedure:
 \begin{enumerate}[noitemsep,nolistsep]
	\item NWChem is used to generate a description of the electronic structure problem in the desired configuration, following~\autoref{sec:review_nwchem}.
	\item We generate a Broombridge representation of C$_{20}$ from the NWChem output following~\autoref{sec:broombridge_generation}.
	\item The C$_{20}$ instance represented in Broombridge is imported, following~\autoref{line:create-from-broombridge} of ~\autoref{lst:specify-hamiltonian}.
	\item Qubitization is chosen as the simulation method, both with and without optimization, following~\autoref{line:chemistry-qubitization} of ~\autoref{lst:obtain-operations}.
	\item A single step of the qubitization walk operator is run through the Trace simulator, which yields resource estimates needed for the simulation, following~\autoref{lst:teleport-host}.
\end{enumerate}

\begin{table}
	\centering
	\begin{tabularx}{\textwidth}{l@{\;\;\;}ll|cccc|ccccc}\hline \hline
		&&&     \multicolumn{4}{|c|}{Generic Qubitization} & \multicolumn{5}{|c}{Optimized Qubitization} \\
		Geometry & $N$ & $\eta$&     Qubits&T gates&$R_z$& $L_1$ norm &      Qubits&T gates&$R_z$ & $L_1$ norm & \autoref{eq:pe-qubitization} $\times$ T gates \\
		\hline
		Ring &  50 & 32 &143&9278072&17931406&1539&312&18605634&18&962& $1.8 \times 10^{13}$ \\
		Fullerene &  50 & 32 &145&18514478&3586938&2004&316&37078492&18&1392&$5.2 \times 10^{13}$ \\
		Bowl &  50 & 32 &145&26153750&36815676&2297&316&52357100&18&1480&$7.7\times 10^{13}$ \\
		Ring &  17 & 14 &63&121574&145222&315&210&252860&18&294&$7.4 \times 10^{10}$ \\
		Fullerene &  17 & 14 &65&242664&292166&448&214&495158&18&418&$2.1 \times 10^{11}$ \\
		Bowl & 17 & 14 &67&357458&569110&411&218&724880&18&384& $2.8 \times 10^{11}$ \\		
		\hline \hline								
	\end{tabularx}
	
	\caption{Resources required to perform a single quantum walk step created by the Qubitization procedure for different configurations of C$_{20}$ in the cc-pVDZ basis set, with $\eta$ electrons and varying numbers of $N$ orbitals within the active space.
	Here all Hamiltonian terms below with norm below $10^{-10}$ Hartree are truncated and the $L_1$ norm is the sum of the absolute values of the Hamiltonian terms in Jordan--Wigner representation.
	Generic qubitization refers to the unoptimized procedure as described in~\autoref{sec:qubitization}, whereas optimized Qubitization applies chemistry-specific optimizations described in~\autoref{sec:optimized_qubitization}.
	The expected number of T gates needed to synthesize $n_{R_z}$ rotations within total error at most $\epsilon$ (without the use of ancill\ae) is bounded above by $3 n_{R_z} \left(\log_2{(n_{R_z}/\epsilon)}+\mathcal{O}(\log\log{(n_{R_z}/\epsilon)})\right)$~\cite{Ross2016Optimal}, where $\Delta=1$ milli\hartree~for chemical accuracy. The rightmost row states the total number of $T$ gates required for an energy estimate of the ground state to chemical accuracy.}
	\label{tab:c20}
\end{table}

The output of the Trace simulator is tabulated in~\autoref{tab:c20}.  
We see from the data included that the costs of implementing the qubitization walk operator is substantially reduced through the use of the optimizations discussed in~\autoref{line:chemistry-qubitization}. 
In particular, for the case where C$_{20}$ is configured in a ring with an active space consisting of $50$ orbitals and using $1$ milli\hartree~as our target precision, we find that if an ancilla-free synthesis method is used then the expected total number of T gates needed to perform the walk operator is roughly $1.8 \times 10^9$.
The majority of this cost arises from synthesis of arbitrary single-qubit rotations with a fault-tolerant gate set.
If we use the optimized approach, then the number of gates required is reduced by nearly a factor of $100$ to $1.8 \times 10^7$ T gates.
The remaining cases see similar improvements, underlining the importance of problem-specific optimization of the subroutines used in qubitization.

As a final example, let us consider the problem of computing the number of gates that would be needed to estimate the ground-state energy within error $\epsilon$ using qubitization.
From~\autoref{eq:pe_cramer-rao}, the number of applications of a unitary required to obtain a phase estimate with standard deviation $\Delta$~\hartree\;is $\approx 1/\Delta$.
Combined with the phase-doubling trick of~\cite{Babbush2018encoding}, the number of applications of the qubitization walk operator is approximately
\begin{equation}
\label{eq:pe-qubitization}
	L_{\exp} \approx \frac{L_1\;\text{norm} }{2\Delta}.
\end{equation}
As the spectrum of the quantum walk is exactly similar to $\arcsin[H/h]\approx H/h$, the only other error contribution is from the finite precision of Hamiltonian coefficients realized in the quantum circuit.
However, we may ignore this error contribution as it can be made arbitrarily small without changing the T-count, to leading order in the optimized qubitization scheme.

By using this formula, we obtain estimates T gate count for sampling from the Hamiltonian spectrum to chemical accuracy.
We find that the number of T gates needed for the simulation (for the ring geometry) is expected to be less than $1.8\times 10^{13}$.
The number of spin orbitals required in this case are $100$, which places the scale on the same order of magnitude as that estimated for Trotter-based simulations for FeMoco given in~\cite{Reiher2016Reaction}.  
Similarly, we find that the number of T gates is within an order of magnitude of the $10^{14}$ gates required for a qualitatively accurate simulation for the $108$ spin-orbital example of FeMoco considered therein.
A crucial difference however is that the FeMoco gate estimates required a certain optimistic extrapolation of empirical Trotter step-size from small, classically simulable molecules.
In contrast, the qubitization T gate estimate is fully rigorous, assuming the ground truth of the presented Hamiltonian.
This further suggests that optimized variants qubitization may be preferable to Trotter--Suzuki simulations for challenging problems in general.

%% file: conclusions.tex

Today, the steady progress of scalable fault-tolerant quantum computing from theory to reality drives intense research in its use.
Many promising quantum algorithms are already known and more continue to be developed at a furious pace.
Paralleling the history of classical computing, we may expect that algorithmic advances will contribute far more to the overall computational capability of quantum computers than optimistic Moore's-Law hardware scaling predictions.
However, the many envisioned applications of quantum computing, particularly that of electronic structure problems, are highly inter-disciplinary.
The current barrier to their use by non-specialists in quantum computing---the intended audience---is, more often than not, insurmountable.
In many cases, effective use of these algorithms requires knowledge of low-level details such as error-correction schemes and specialized circuit optimizations.

When the requisite quantum hardware arrives, its use should be accessible, reproducible, and extensible. 
The open-source tools and workflow we present here are intended to realize this vision.
Accessibility by the target audience is achieved by using NWChem, a standard classical computational chemistry packages, as the access point, followed by the integration with quantum algorithms that are expressed and invoked at a high-level in our chemistry library. 
Reproducibility through our proposed workflow allows for a straightforward and consistent application to future problems of scientific and industrial interest.
Extensibility of our work to future algorithmic improvements on both the classical and quantum side remains possible through our definition of standardized interfaces, such as Broombridge, for representing electronic structure problems.

The examples we provide and their results underscore the value of the workflow enabled by integrating \qsharp~with NWChem.
By combining the two, we are capable of generating end-to-end resource estimates for simulation that are not only accurate but are also highly optimized and reproducible.
This illustrates that by building libraries to compute accurate electronic structure representations and also by building highly optimized libraries for quantum simulation targeted at quantum hardware, we can create tools that empower people to explore and cost quantum chemistry simulation algorithms without requiring that the user be an expert in both fields.

This work highlights the impact that scientific software development can have on reproducible research.
It is our hope that this work helps chemists, physicists, and computer scientists to pool their knowledge so as to enable quantum  methods for electronic structure calculation to reach the same level of sophistication and value as classical computing today already achieves.
More broadly, we hope to inspire the development of new libraries, platforms, and even programming languages that facilitate the entry of new researchers into the interdisciplinary field of quantum computing and eventually be constructively used to solve the intractable quantum problems of today.

%% file: apx-running-examples.tex
\lstMakeShortInline[language=Bash]|

The code examples in this paper are available in the \lstinline+anc/src/+ directory of the supplementary material. 
Additional examples may be found at in samples repository of Microsoft Quantum Development Kit at~\url{https://github.com/Microsoft/Quantum}.
The implementation of the Microsoft Quantum Development Kit chemistry library may be found at~\url{https://github.com/Microsoft/QuantumLibraries}.

To run the examples on your own system, we recommend either installing the Microsoft Quantum Development Kit and NWChem, or using the provided Docker file to prepare a container for use with these examples.
We briefly describe both approaches in this appendix.

We also provide a configuration file for Docker, a software platform for managing lightweight computing environments known as \emph{containers}.
This \lstinline+Dockerfile+ can be used to automatically build and run a container that includes the Microsoft Quantum Development Kit, NWChem, and the code examples in this paper.
Though installing Docker is beyond the scope of this paper, complete instructions can be found at \url{https://docs.docker.com/install/}.

Once Docker has been installed on your machine, the |docker| command can be used to automatically build and run the container described by \lstinline+anc/src/Dockerfile+, as shown in \autoref{lst:docker-build}.
We format the procedure in \autoref{lst:docker-build} for use with Bash or similar shells, but similar instructions hold for PowerShell and other command-line environments.

\begin{lstlisting}[
    language=Bash,
    label={lst:docker-build},
    caption={Procedure to run the teleport example of \autoref{lst:teleport} using Docker.\hfill}
]
    # First, navigate to the //*src*// folder of the supplementary material.
    $ cd anc/src
    # This command will cause Docker to build a container called //*chem-examples*//
    # from a Dockerfile found in the current working directory (denoted //*.*//).
    # Note that this command needs to download large images, and may take a long
    # time to complete. These images are then cached, such that subsequent builds
    # will complete more quickly.
    $ docker build -t chem-examples .
    # Once this completes, the new container can be run in an interactive mode.
    $ docker run -it chem-examples
    # This will start a PowerShell prompt inside the container, appropriate for
    # use with commands such as those discussed in //*\autoref{sec:broombridge_generation}*//.
    PS /src> cd examples/teleport
    PS /src/examples/teleport> dotnet run
    Used 2 CNOT operations.
\end{lstlisting}

\lstDeleteShortInline|